\documentclass[aps,prl,reprint,amssymb,showpacs,superscriptaddress,notitlepage,onecolumn]{revtex4-2}
\usepackage{bm}
\usepackage{graphicx,hyperref}
\usepackage{dcolumn}
\usepackage{braket}
\usepackage{natbib}
\usepackage{amsmath}
\usepackage{color}
\usepackage{mathtools,amssymb}
\usepackage{soul}

\usepackage{float}
\usepackage{amssymb}
\usepackage{esint}
\usepackage{mathtools}
\usepackage{physics}
\usepackage{makecell}
\usepackage{multirow}
\usepackage{array}
\usepackage{geometry}
\usepackage{url}

\begin{document}
\title{Sub-wavelength optical lattice in 2D materials}

\author{Supratik Sarkar}
\altaffiliation{These authors contributed equally to this work}
\affiliation{Joint Quantum Institute, University of Maryland, College Park, MD 20742, USA}

\author{Mahmoud Jalali Mehrabad}
\altaffiliation{These authors contributed equally to this work}
\affiliation{Joint Quantum Institute, University of Maryland, College Park, MD 20742, USA}
\email{mjalalim@umd.edu}

\author{Daniel G. Suárez-Forero}
\altaffiliation{These authors contributed equally to this work}
\affiliation{Joint Quantum Institute (JQI), University of Maryland, College Park, MD 20742, USA}

\author{Liuxin Gu}
\altaffiliation{These authors contributed equally to this work}
\affiliation{Department of Materials Science and Engineering, University of Maryland, College Park, MD 20742, USA}
\affiliation{Maryland Quantum Materials Center, College Park, Maryland 20742, USA}

\author{Christopher J. Flower}
\affiliation{Joint Quantum Institute, University of Maryland, College Park, MD 20742, USA}

\author{Lida Xu}
\affiliation{Joint Quantum Institute, University of Maryland, College Park, MD 20742, USA}

\author{Kenji Watanabe}
\affiliation{Research Center for Electronic and Optical Materials, National Institute for Materials Science, 1-1 Namiki, Tsukuba 305-0044, Japan}

\author{Takashi Taniguchi}
\affiliation{Research Center for Materials Nanoarchitectonics, National Institute for Materials Science,  1-1 Namiki, Tsukuba 305-0044, Japan}

\author{Suji Park}
\affiliation{Center for Functional Nanomaterials, Brookhaven National Laboratory, Upton, NY 11973, USA}

\author{Houk Jang}
\affiliation{Center for Functional Nanomaterials, Brookhaven National Laboratory, Upton, NY 11973, USA}

\author{You Zhou}
\email{youzhou@umd.edu}
\affiliation{Department of Materials Science and Engineering, University of Maryland, College Park, MD 20742, USA}
\affiliation{Maryland Quantum Materials Center, College Park, Maryland 20742, USA}

\author{Mohammad Hafezi}
\email{hafezi@umd.edu}
\affiliation{Joint Quantum Institute, University of Maryland, College Park, MD 20742, USA}


\begin{abstract}
\textbf{Recently, light-matter interaction has been vastly expanded as a control tool for inducing and enhancing many emergent non-equilibrium phenomena. However, conventional schemes for exploring such light-induced phenomena rely on uniform and diffraction-limited free-space optics, which limits the spatial resolution and the efficiency of light-matter interaction. Here, we overcome these challenges using metasurface plasmon polaritons (MPPs) to form a sub-wavelength optical lattice. Specifically, we report a ``non-local" pump-probe scheme where MPPs are excited to induce a spatially modulated AC Stark shift for excitons in a monolayer of MoSe$_2$, several microns away from the illumination spot. Remarkably, we identify nearly two orders of magnitude reduction for the required modulation power compared to the free-space optical illumination counterpart. Moreover, we demonstrate a broadening of the excitons' linewidth as a robust signature of MPP-induced periodic sub-diffraction modulation. Our results will allow exploring power-efficient light-induced lattice phenomena below the diffraction limit in active chip-compatible MPP architectures.}

\end{abstract}


\maketitle

\subsection{Introduction}

Recent decades have witnessed a surge of interest in using light as a powerful tool to induce, control, and enhance exotic emergent non-equilibrium phenomena in quantum and topological materials \cite{de2021colloquium,bloch2022strongly}. For example, light has been widely utilized for inducing AC Stark shift \cite{sie2015valley,kim2014ultrafast,uto2024interaction,evrard2024nonlinear,cunningham2019resonant,lamountain2021valley}, Floquet bands \cite{wang2013observation,rudner2020band}, signatures of magnetization \cite{wang2022light,disa2023photo}, superconductivity \cite{rowe2023resonant}, anomalous Hall effect \cite{mciver2020light}, and melting gapped state into semi-metallic ones \cite{huber2024ultrafast}. Common schemes for exploring such light-induced phenomena are based on diffraction-limited free-space optical spectroscopic techniques, which do not provide spatial patterns to imprint sub-wavelength lattice features \cite{kim2020optical}. These methods fundamentally put a bound on the spatial resolution and the efficiency of light-matter interaction due to the limited confinement of free-space optical beams. Moreover, pump-probe spectroscopy of emergent light-induced phenomena typically is performed ``locally", i.e., the pump and the probe illuminate and investigate the same spot. Such local pump-probe techniques cause unavoidable parasitic thermal and pump-induced noise, which necessitates complex heat management and pump-rejection methods \cite{maly2017polarization}. The development of innovative platforms and spectroscopy techniques to address these challenges remains an active area of research \cite{zhang2022nano}.

Concurrently, the emergence of emitter-integrated nano-plasmonics has enabled strong sub-wavelength confinement of optical fields for efficient light-matter interaction. Of particular interest are localized and propagating surface plasmon polaritons (SPPs), which are excitations at the metal-dielectric interface whose electromagnetic field decays exponentially with distance from the surface, offering remarkable control at near-field \cite{gramotnev2010plasmonics,zheludev2022optical,basov2016polaritons}. For example, active nano-plasmonic devices have been recently used for strong light-matter coupling \cite{lee2017electrical,zhang2023observation,wang2016coherent,niu2022unified,zheng2017manipulating}, probing dark excitons and cascaded energy transfer \cite{zhou2017probing,shi2017cascaded} and directional emission \cite{li2023versatile}. On the downside, these devices offer limited propagation length and functionalities, which motivated the development of metasurface plasmon polaritons (MPPs) \cite{poddubny2013hyperbolic,high2015visible,gomez2015hyperbolic} with superior performance. MPPs have enabled several interesting effects such as Purcell-enabled optical modulators \cite{li2023purcell}, exciton manipulation \cite{sun2019separation}, enhanced harmonic generation \cite{shi2018plasmonic}, and plasmon-exciton strong coupling \cite{yu2021quantifying,zhang2023strong}, which are explored with local optical spectroscopy.

Here, we design and realize an MPP-based non-local pump-probe scheme for inducing sub-wavelength optical lattices. Specifically, the excited MPPs imprint a one-dimensional lattice for excitons, resulting from a spatially modulated AC Stark shift.


\begin{figure}
    \centering
    \includegraphics[width=\columnwidth]{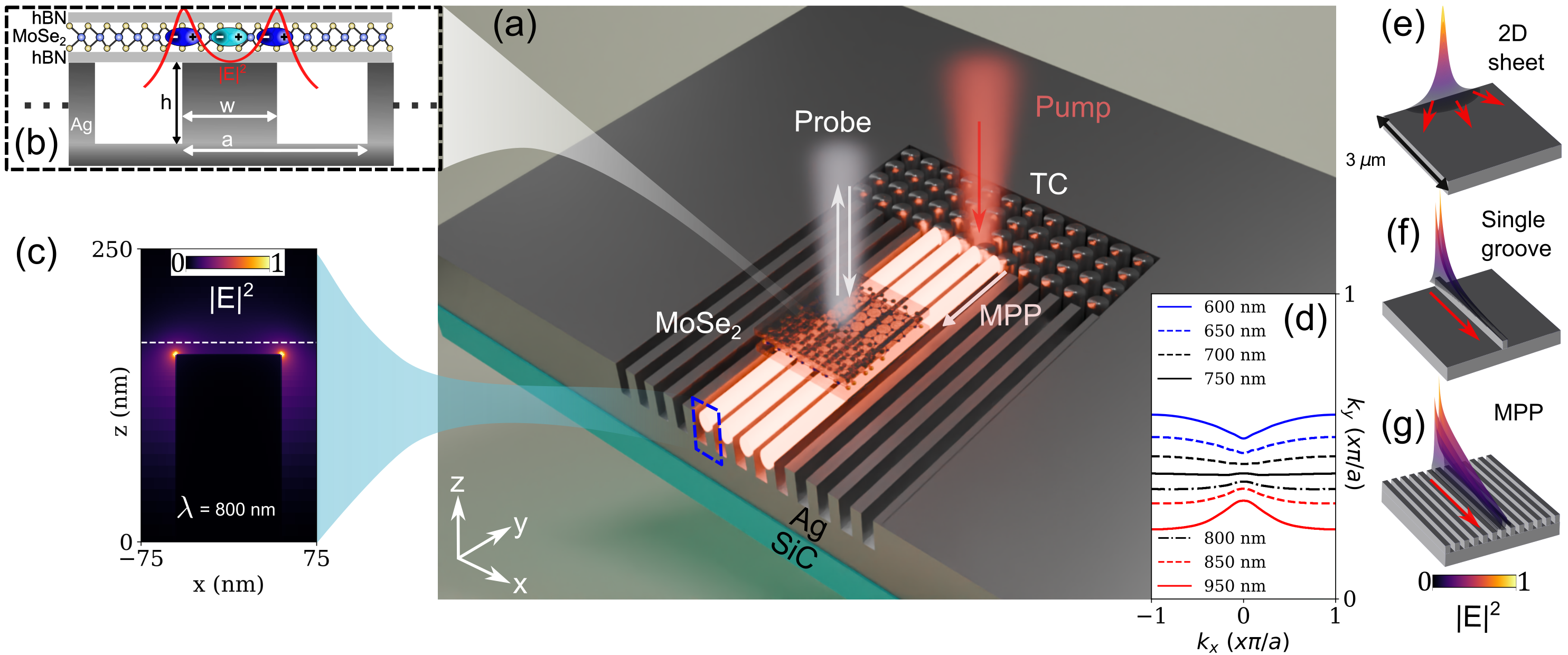}
    \caption{\textbf{Sub-diffraction nonlocal pump-probe spectroscopy of excitons.} (a) Scheme for inducing AC Stark shift on a MoSe$_2$ monolayer using metasurface plasmon polaritons (MPPs) in a 1D lattice of silver grooves. The periodically modulated MPP modes are excited with a pump laser (red beam) at the top coupler (TC). The MPP device is designed to strongly confine the plasmonic modes, which propagate in the y-direction of the lattice with negligible diffraction within $700-800$ nm, which is close to the excitonic resonance of $\sim 750$ nm in MoSe$_2$. A secondary broadband laser (white beam) probes the MoSe$_2$ excitons. (b) Schematic of the device cross-section, including the hBN-encapsulated MoSe$_2$ on the silver grooves. The periodicity of the lattice is $a=150$ nm, the height is $h=160$ nm, and the width is $w=90$ nm. The red curve shows the strong $|E|^2$ modulation at the position of the MoSe$_2$ monolayer for a pump of 800 nm. Excitons present at positions of peak electric field intensity observe larger blue shifts compared to the ones near the center of the groove. (c) Normalized electric field intensity $|E|^2$ at the x-z cross-section of the device. The position of the TMD is marked by the dashed horizontal line, which is 10 nm above the surface.
    (d) Iso-frequency contours of the MPP device for different wavelengths of MPPs showing suppressed diffraction in the range 700-800 nm. (e-g) Finite-difference time-domain (FDTD) simulations comparing SPP propagation loss over a distance of 3 $\mu$m on a 2D sheet of silver, a single groove waveguide, and the proposed MPP device at 800 nm respectively. In (e-g) a dipole source placed 10 nm above the surface is used to excite the SPPs.}
    \label{schematic}
\end{figure}

\subsection{Concept}

Figure~\ref{schematic}(a) illustrates our scheme. The metasurface device consists of a 1D array of silver nano-grooves, and is designed to support a 1D lattice of MPPs with negligible diffraction around wavelengths close to the excitonic resonance. A pump laser (marked in red) is coupled into the MPP modes through the top coupler. The traveling MPPs form a sub-diffraction periodic potential for the MoSe$_2$ monolayer, whose exciton resonance lies at a wavelength $\sim750$ nm. A broadband beam (marked in white) probes the modulation of the transition metal dichalcogenide (TMD) excitons at a spot distant from the top coupler.  

Figure~\ref{schematic}(c) shows the normalized electric field intensity $|E|^2$ in the x-z plane (shown only for one groove) at 800 nm obtained from finite-difference time-domain (FDTD) simulations. The electric field intensity is maximum at hot spots near the sharp corners of the air-silver interface and decays exponentially with distance from the surface; this is a characteristic feature of SPPs. The TMD is located $\sim10$ nm above the surface, as marked by the dashed white line in Fig.~\ref{schematic}(b). The red curve shows the electric field intensity $|E|^2$ at the monolayer's position. Importantly, due to this sub-diffraction variation of $|E|^2$, the excitons experience a position-dependent AC stark shift: those located at the local maxima of $|E|^2$ experience a larger blue shift compared to other excitons (represented by the dark and light blue ellipses in Fig.~\ref{schematic}(b); see Supplementary Information (SI) for detailed study of spatial dependence of $|E|^2$ and the full scheme of the experimental setup).

The diffractionless propagation of MPPs is a fundamental advantage for our non-local AC Stark shift scheme. To understand this feature, we calculate the dispersion of the device using the FDTD method. Figure~\ref{schematic}(d) shows the iso-frequency contours of the device for different wavelengths. Here, $k_x$ and $k_y$ are the wavevectors along the x and y directions. Notably, at wavelengths of $\sim 700-800$ nm, the MPPs propagate with maximally suppressed diffraction. Moreover, for lower wavelengths (blue lines) the dispersion is anisotropic and hyperbolic with a negative angle of refraction, whereas it is elliptical with a positive angle of refraction for longer wavelengths (red line). Our numerical simulations closely follow the analytical dispersion relationship of the MPP device given by the coupled waveguide formalism under a slowly-varying amplitude approximation \cite{eisenberg2000diffraction, fan2006all}:
$k_y = k_{wg} + 2t\cos(k_x a)$, where $k_{wg}$ is the propagation constant of the MPP waveguide lattice, and $t$ is the tunneling amplitude between nearest-neighbor grooves in the lattice (see SI for details). This behavior is analogous to the recent observation of canalized polaritons originating from flat iso-frequency contours in twisted MoO$_3$ heterostructures \cite{duan2023multiple,obst2023terahertz}.

Moreover, the choice of an MPP device structure over other conventional plasmonic platforms such as 2D sheets and single grooves provides a noticeable advantage in long-distance power delivery due to their diffractionless propagation. To confirm this, the propagation of the excitation for each case is shown in Fig.~\ref{schematic}(e-g). Over a propagation length of 3 $\mu$m, the MPP device has $\sim 5$ and $\sim 50$ folds enhanced power transmission compared to its single groove, and 2D sheet counterparts, respectively. Details of the simulations and a more comprehensive analysis of the wavelength-dependent operation of our device can be found in the SI.


\subsection{Device fabrication and characterization}

\begin{figure}
    \centering    
    \includegraphics[width=\columnwidth]{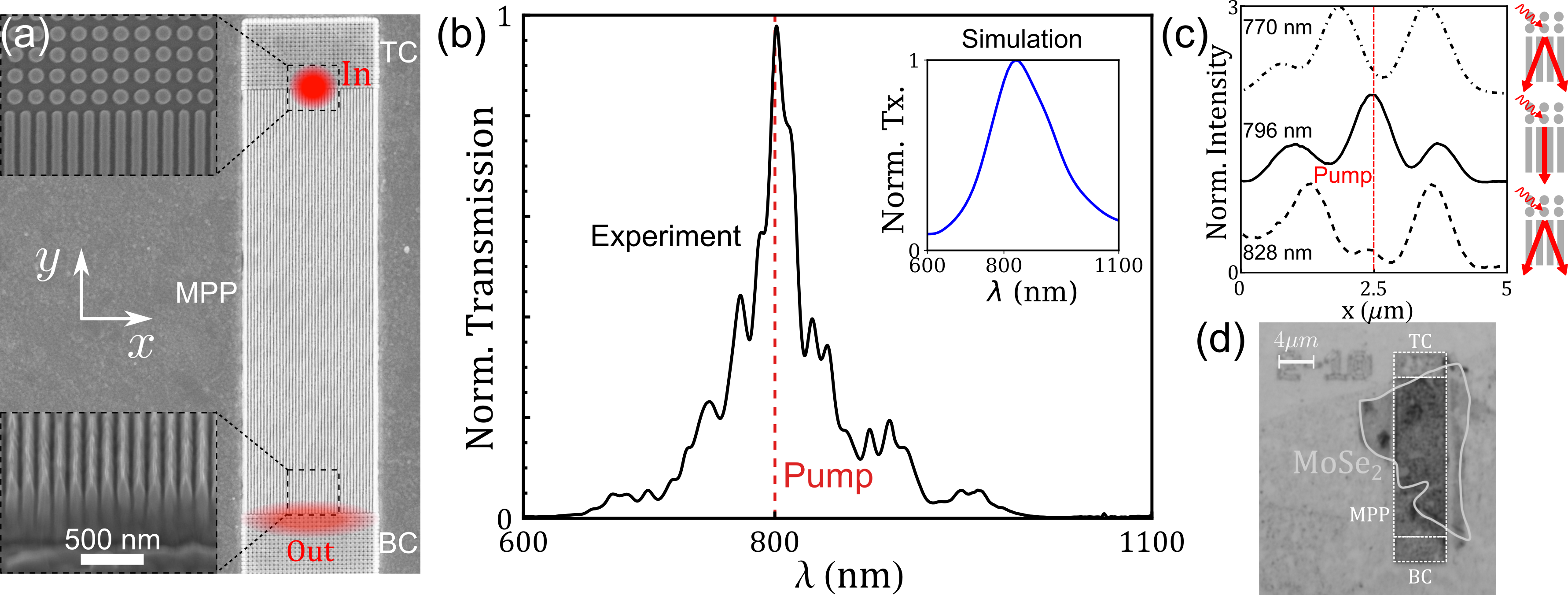}
    \caption{\textbf{MPP device fabrication and characterization.} (a) High-resolution scanning electron microscope (SEM) images of representative fabricated MPP device. A 1D array of grooves is terminated with a top coupler (TC) and a bottom coupler (BC) to in-couple and out-couple the propagating MPPs. The top left inset shows a close-up view of the TC and the grooves. The bottom left inset shows a perspective view with a cross-section of the grooves. (b) The measured transmission spectrum (normalized by the broadband input laser spectrum) of the MPP device, shows a transmission peak around the pump wavelength of 800 nm. The inset shows the simulated transmission spectrum. (c) The spatial pattern of the emission from the output coupler for three different wavelengths: 770 nm, 796 nm and 828 nm. The red dashed line shows the position of the pump laser. (d) A false-color microscope image of the device after stacking the MoSe$_2$ monolayer. TC, BC and the MPP device are marked by dashed boxes.}
    \label{device}
\end{figure}

Figure \ref{device}(a) shows a high-resolution scanning electron microscopy (SEM) image of a representative MPP device, which is fabricated on a 500 nm thick silver layer sputtered on a silicon carbide substrate. Each device consists of 40 silver grooves (each 20 $\mu$m long) and has a top coupler, and a bottom coupler on each end of the metasurface. The top and bottom left insets show the top coupler-MPP device interface, and a cross-section of the waveguides, respectively. To prevent the silver from oxidation, the chip is coated with a 5 nm thick layer of Al$_2$O$_3$. Further details of the fabrication are available in the SI.



We begin by characterizing the transmission spectrum of the device before stacking the TMD on the top, by coupling a broadband laser into one of the couplers and measuring the output from the other. The results are shown in Fig.~\ref{device}(b).
The transmission is maximum for a wavelength of $\sim800$ nm and it decreases as a function of the detuning, in agreement with the simulated iso-frequency contours (Fig.~\ref{schematic}(d)). The inset shows the simulated normalized transmission spectrum for the device monitored after 3 $\mu$m propagation of the MPPs. At greater detuning from the peak at $\sim800$ nm, the measured transmission is smaller than in the simulation, since at large diffraction angles the MPPs do not successfully reach the bottom coupler. The asymmetric nature of the normalized transmission can be attributed to the longer propagation lengths for the elliptical MPPs \cite{high2015visible}.
We further note the observation of oscillations in the transmission spectra, possibly originating from the confinement of light by the ends of the 20 $\mu$m long MPP device forming a Fabry-Perot mode.
To experimentally characterize the diffraction in the MPP device, we measure the spatial pattern of the emission from the output coupler for three different wavelengths as shown in Fig.~\ref{device}(c). The red line shows the position of the pump laser at the input coupler. We observe diffracted outputs at wavelengths of 770 nm and 828 nm, whereas there is minimal diffraction at 796 nm, which is very close to our designed pump wavelength. Simulation and details of this wavelength-dependent behavior are available in the SI. After the initial MPP device characterizations, we complete the integrated device by stacking the hBN-encapsulated MoSe$_2$ monolayer on the MPP device. Figure~\ref{device}(d) shows a false-color microscope image after the stacking procedure.


\subsection{Nonlocal and local AC Stark shift measurements}

\begin{figure}
    \centering
    \includegraphics[width=0.85\columnwidth]{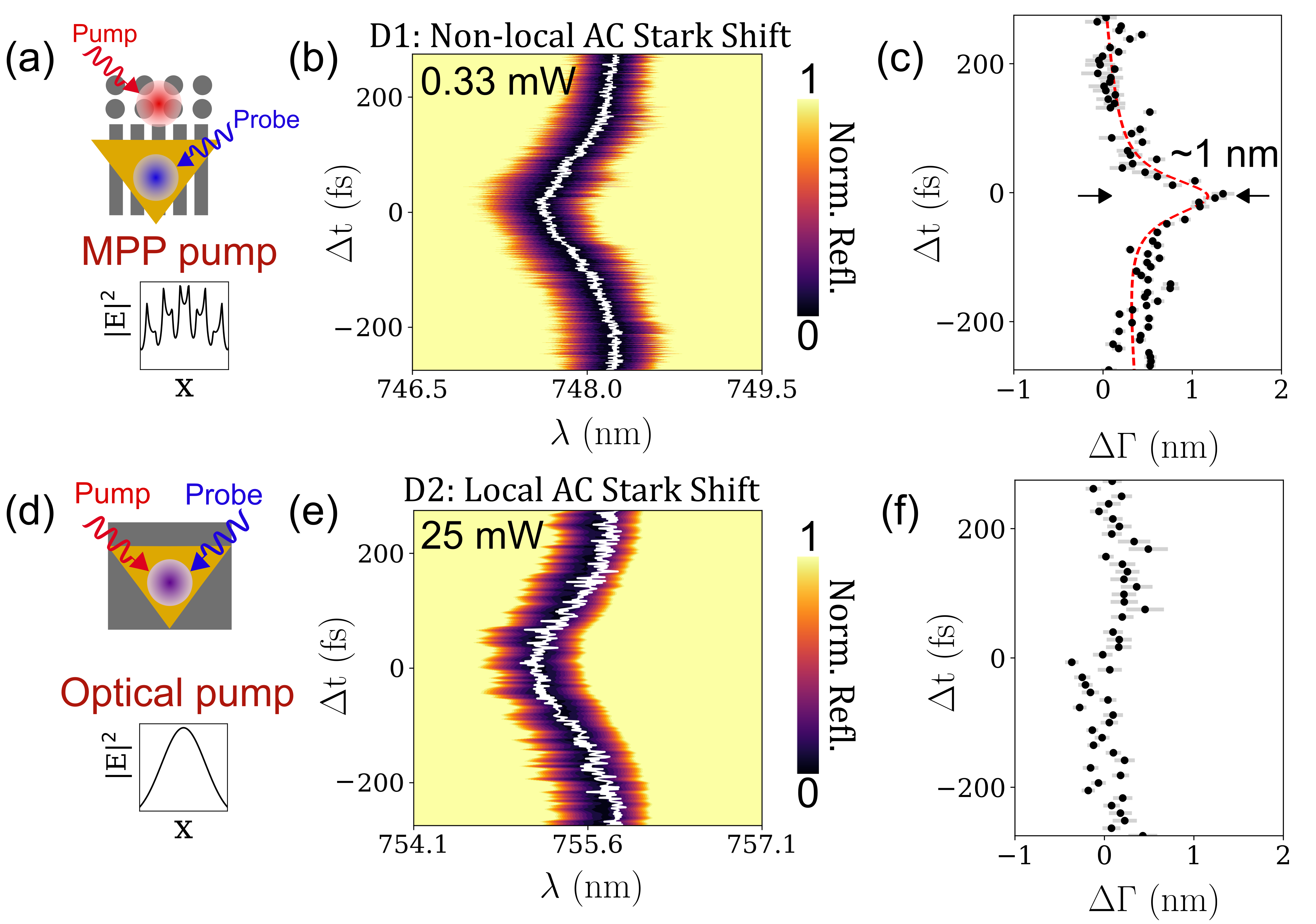}
    \caption{\textbf{Nonlocal modification of exciton energy with MPPs.} (a) Schematic of our nonlocal AC Stark shift measurement scheme that utilizes a sub-diffraction spatially-modulated MPP pump. The device is pumped at the top coupler to generate MPP modes in the metasurface. The MoSe$_2$ is probed $\sim 3 \mu$m away from the pump. (b) Nonlocal AC Stark shift was measured in device D1 while the delay between the pump and probe pulses was varied. The white lines show the extracted exciton peak wavelength from numerical fitting. (c) The extracted change in linewidth $\Delta \Gamma$ of the exciton extracted from the numerical fitting of the nonlocal AC Stark shift data.
    (d) Schematic of traditional local AC Stark shift measurement scheme that utilizes a diffraction-limited Gaussian optical pump.
    (e) Local AC Stark shift measured in device D2 (MoSe$_2$ on silver substrate).
    (f) The extracted change in linewidth $\Delta \Gamma$ of the exciton extracted from the numerical fitting of the local AC Stark shift data. The error bars represent one standard deviation error in parameter estimation from numerical fitting.
    }
    \label{stark}
\end{figure}

To demonstrate the sub-diffraction non-local AC Stark shift scheme introduced in Fig.~\ref{schematic}, we excite the MPP modes using a 150 fs pulsed laser at 800 nm wavelength, as shown schematically in Fig.~\ref{stark}(a). The pump excites the MPPs in several grooves that lie under our diffraction-limited pump (see the SI for a discussion on the multi-groove excitation). They travel along the y-direction of the device and modulate the MoSe$_2$ monolayer located on top of the MPP device. Importantly, the excited MPPs have a spatial periodic intensity profile with sharp features below the diffraction limit (see SI for simulations of the mode profiles). To probe the AC stark shift induced by the MPPs, we perform absorption spectroscopy using a weak pulsed broadband laser placed 3 $\mu$m away from the pump in a reflection configuration. We note that the pump is red-detuned by $\sim 50$ nm from the excitonic energy to prevent photoluminescence.

Figure~\ref{stark}(b) shows the normalized reflectivity spectrum as a function of pump-probe temporal delay $\Delta t$. Positive $\Delta$t refers to the probe reaching the sample before the pump, and vice-versa. We observe an AC Stark shift of $\sim$0.65 nm at $\Delta t=0$. The white line shows the extracted exciton wavelength from the numerical fitting (see the SI for the detail of the numerical fitting). In the data presented in Fig.~\ref{stark}(b), we pump the top coupler with 100 mW of power to create the MPPs. Taking into account the estimated efficiency of the device, this translates to an ``MPP pump" power of 0.33 mW at the position of the TMD under the probe. The complete analysis of the power efficiency, as well as pump power and pump detuning dependence of the AC Stark shift, is presented in the SI.


\subsubsection{Power efficiency comparison}

Next, we confirm the superior power efficiency of our non-local pump-probe scheme, which arises from MPP-induced strong field confinement, by considering the optical power required to generate the same AC Stark shift using the conventional local pump-probe counterpart. In the local case, the pump and the probe are spatially overlapped as shown in Fig.~\ref{stark}(d). We perform the local AC Stark shift measurement on a TMD placed on an unpatterned silver substrate (device D2) as shown in Fig.~\ref{stark}(e). We observe an identical AC shift of $\sim0.65$ nm using $\sim25$ mW of pump power. Remarkably, when compared to the MPP-induced non-local AC Stark shift with an effective pump power of 0.33 mW, the non-local AC Stark shift scheme requires $\sim 90$ times less pump power to observe a comparable AC Stark shift for the same detuning (see SI for details). Numerically, we estimate that the maximum field enhancement at the position of the TMD for our non-local scheme can be as high as 130 times the diffraction-limited optical pump intensity (see SI for detailed calculation). The slight discrepancy between the experimental observation and numerical estimate is likely due to fabrication imperfections like surface roughness, rounded groove corners, and deviation of the TMD position from its nominal value in the z-direction. Moreover, this remarkable observed enhancement is further substantiated by measuring a similar non-local AC Stark shift effect on a different device, which highlights the robustness of our scheme (see the data for device D4 in SI). Furthermore, we confirm that the MPP-enabled strong field confinement is central to our scheme compared to 2D sheets (discussed earlier in Fig.~\ref{schematic}(e-g)). To achieve this, we perform the same non-local scheme on a similar TMD device placed on a 2D sheet. Using an identical pump power strength, no AC Stark shift is observed in this case (see the non-local measurement of device D2 in SI).


\subsubsection{Sub-diffraction modulation of excitons}

We demonstrate another intriguing optical lattice-induced effect, in the form of an excitonic linewidth broadening, presented in Fig.~\ref{stark}(c). The excitons under our diffraction-limited probe laser experience a strong spatial variation of the electric field intensity of the MPPs as shown in Fig.~\ref{stark}(a). 
This periodic sub-diffraction modulation of electric field intensity (which can vary as strongly as $\sim 50$\% as shown in the inset of Fig.~\ref{stark}(a)) maps to a periodic modification of AC Stark shift of excitons. Excitons located at the maximum of the electric field intensity experience greater blue shift compared to excitons in their vicinity. This leads to inhomogeneous AC Stark shift of the excitons, thereby resulting in a linewidth broadening of $\sim 1$ nm at $\Delta t=0$. A theoretical model for the expected linewidth broadening resulting from the spatial profile of the electric field intensity is provided in SI and closely matches our experimental observations.
This observation is in sharp contrast with the local counterpart, where there is no noticeable change in the linewidth at $\Delta t=0$ (Fig.~\ref{stark}(f)) (for a similar observation and discussion for a TMD monolayer placed on an unpatterned silica substrate, see the data for the device D3 in SI).


\subsection{Summary and outlook}

In summary, we demonstrated a TMD-integrated plasmonic metasurface approach for efficient modulation of excitons below the diffraction limit. Due to its intrinsic non-local pump-probe character \cite{pizzuto2021nonlocal}, our platform provides a useful protocol for spatial on-chip pump rejection as well as suppressing parasitic thermal effects \cite{maly2017polarization}.
More generally, our results suggest that lattice SPPs in MPP device architectures can become an innovative and versatile toolbox to actively manipulate matter excitations below the diffraction limit.

Beyond our exemplary AC Stark shift case, a wide range of other light-induced phenomena can be investigated below the diffraction limit with superior power efficiency and pump noise suppression. This includes light-induced magnetization \cite{wang2022light}, superconductivity \cite{dehghani2021light}, quantum Hall effects \cite{mciver2020light}, electronic polarization \cite{zhang2024light}, polariton modulators \cite{lee2024ultra} and plasmonic effects in van der Waals heterostructures \cite{kipp2024cavity,tabataba2024metasurface}. Another intriguing direction is the sub-wavelength periodic modulation of materials and matter excitations harnessing the on-demand spectral and spatial tunability of our MPPs. An example includes optical imprinting of lattices \cite{kim2020optical}, where the optical drive forms a lattice structure in space. The lattice nature of MPPs also may enable engineering chiral light-matter interaction \cite{sun2019separation} using plasmonic spin Hall states \cite{high2015visible}. Moreover, the sub-diffraction spatial modulation and the electric field confinement can be probed using near-field optical microscopy for a better understanding of the MPP properties, and allow further innovative integration with solid-state emitters \cite{frischwasser2021real,vincent2021scanning,betzig1992near,zhang2022nano}.

In addition to light-induced phenomena, our platform could be useful for cavity-induced physics \cite{bloch2022strongly,schlawin2022cavity,garcia2021manipulating,basov2016polaritons}. In particular, the strong spatial confinement of light in such plasmonic structures combined with the hyperbolic photonic modes of the lattice can lead to intriguing phenomena such as meditated long-range interactions between distant emitters \cite{cortes2017super,boddeti2024reducing,narimanov2024hyperbolic}.


\subsection{Methods}

\textbf{Device fabrication}:
The single-crystalline silver films (500 nm thick) and buffer layer platinum film were deposited by DC sputtering on a SiC wafer, similar to the process reported in Ref.~\cite{high2015visible}. Next, the devices were patterned on the silver by electron beam lithography with an Al$_2$O$_3$ hard mask, followed by reactive ion etching and plasma etching. After etching, another thin layer of Al$_2$O$_3$ is deposited on the devices to prevent the silver from oxidizing.
After initial characterization of the plasmonic devices, the hBN-MoSe$_2$-hBN heterostructure was transferred onto the device using the standard dry transfer method with PDMS (Polydimethylsiloxane) and PC (Polycarbonate) stamps. Complete details of the device fabrication are available in the SI.

\textbf{Experimental setup}:
The sample was mounted in a Montana cryostat at 4 K. For reflectivity measurements, we used a confocal microscopy setup. A Ti:Sapphire laser tuned at 800 nm, with a pulse duration of 150 fs and a repetition rate of 80 MHz was used as the pump. The pump beam was split with a polarizing beam splitter, and one branch of it was sent into a supercontinuum generation non-linear fiber to create a broadband probe pulse. Meanwhile, the other branch was sent to a motorized delay stage. This allowed us to control the relative path length difference between the pump and the probe pulses up to a precision of 100 nm. Next, the pump and the generated broadband probe beams were sent through an optical chopper operating at 1 kHz. This was to prevent heating the sample while having a high pump peak intensity. The chopper was placed at the focal spot of a telescope such that both pump and probe were focused on the same spot at the chopper. A half-waveplate linear polarizer combination was used to cross-polarize the residual pump before detection of the signal in a spectrometer with a 300 grooves per mm diffraction grating, and a CCD camera. A detailed schematic of the experimental setup is available in the SI.

\textbf{FDTD simulations}:
The electromagnetic field profiles and iso-frequency contour simulations were performed using the FDTD method in Lumerical. A dipole placed 10 nm above the surface was used to excite the MPP modes of interest in the simulations, unless otherwise specified. The propagating MPPs were monitored using frequency domain monitors to extract quantities like propagation loss, coupling efficiency to gratings, field confinement, far-field radiation pattern, etc. The iso-frequency contours were calculated by performing Fourier transformations of the simulated field profiles in the x-y plane over a large area.


\subsection{Acknowledgements}

The authors wish to acknowledge fruitful discussions with Aaron Sternbach, Zubin Jacob, Andrey Grankin, Deric Session, and Nick Martin. 

\subsection{Funding}

S.S., M.J.M., D.G.S.F., C.F., L.X., M.H. were supported by ARO W911NF2510066, DARPA HR00112490310 and HR00112530313. L.G. and Y.Z. were supported by ARO W911NF2510066, NSF DMR-2145712, and the Department of Energy (DOE) DE-SC-0022885 grants. This research used Quantum Material Press (QPress) of the Center for Functional Nanomaterials (CFN), which is a US Department of Energy, Office of Science User Facility, at Brookhaven National Laboratory under contract no. DE-SC0012704. K.W. and T.T. acknowledge support from the JSPS KAKENHI (grant nos. 20H00354, 21H05233 and 23H02052) and World Premier International Research Center Initiative (WPI), MEXT, Japan, for hBN synthesis.




\subsection{Competing interests}
The authors declare no competing interests.


\subsection{Data availability}
All of the data that support the findings of this study are reported in the main text and Supplementary Information. Source data are available from the corresponding authors on reasonable request.


\bibliography{Main.bib}
\newpage

\newpage


\setcounter{equation}{0}
\setcounter{figure}{0}
\setcounter{table}{0}
\setcounter{page}{1}
\makeatletter
\renewcommand{\theequation}{S\arabic{equation}}
\renewcommand{\thefigure}{S\arabic{figure}}
\pagenumbering{roman}

\subsection{\Large Supplementary Information}


\subsection{Contents}

\noindent {\bf \hyperref[si:fab]{1. Device fabrication}}
\\{\bf \hyperref[si:setup]{2. Experimental setup}}
\\{\bf \hyperref[si:wavelength]{3. Wavelength-dependent behavior of plasmonic metasurface}}
\\{\bf \hyperref[si:IFC_duty_cycle]{4. Variation of dispersion with duty cycle}}
\\{\bf \hyperref[si:theory]{5. Coupled mode theory}}
\\{\bf \hyperref[si:budget]{6. Power budget and field confinement}}
\\{\bf \hyperref[si:multiple_dipoles]{7. Exciting multiple waveguides in the metasurface}}
\\{\bf\hyperref[si:defect]{8. Exciting a single waveguide in the metasurface}}
\\{\bf \hyperref[si:control]{9. Control experiments}}
\\{\bf \hyperref[si:fitting]{10. Data processing and fitting routine}}
\\{\bf \hyperref[si:power_broadening]{11. Power broadening estimation}}
\\{\bf \hyperref[si:shift_broad]{12. MPP-induced linewidth broadening - AC Stark shift relationship
}}

\newpage

\subsection{\label{si:fab} 1. Device fabrication}

\subsubsection{Single-crystalline silver films deposition}

The single-crystalline silver films and buffer layer platinum film were deposited by DC sputtering (AJA International 1800) on a SiC wafer, similar to the process reported in Ref. \cite{high2015visible}. Prior to the deposition, the SiC chip was first cleaned by sonicating in acetone and isopropyl alcohol respectively for 5 minutes each, followed by soaking in Piranha solution (3:1 sulfuric acid: hydrogen peroxide) for 5 minutes to remove all organic residuals. Next, the chip was soaked in 49\% hydrofluoric acid for 3 minutes and then put into boiled DI water for 10 minutes to remove any residual chemical bonds on the surface. Then, the chip was transferred to a sputtering chamber immediately. 
The base pressure for the sputtering deposition was $1\times 10^{-7}$ Torr. The deposition was done at room temperature (25$^{\circ}$C). The deposited thickness of Pt and Ag film (Pt 100 nm, Ag 500 nm) was calculated based on the deposition rate.

\subsubsection{Plasmonic metasurface fabrication process}

After the silver deposition, 5 nm of Al$_2$O$_3$ hard mask was deposited on top of the chip via the atomic layer deposition (ALD) method at 90$^{\circ}$C. The thickness was confirmed with a Woollam Spectroscopic Ellipsometer. Then we performed electron beam lithography on the chip using 1:1 ZEP-520A:Anisole as the resist. Next, we grow 90 nm Al$_2$O$_3$ on top by ALD at 110$^{\circ}$C on the developed device. The hard mask layer was then dry etched by RIE etching with a mixture of BCl$_3$ and Cl$_2$ gases. After etching off the alumina, the chip was soaked in remover PG for 40 minutes to remove the resist residue. Then, all the patterns were transferred onto the underlying 5 nm Al$_2$O$_3$ mask.
We then transferred the pattern from the Al$_2$O$_3$ hard mask to the Ag film underneath by a high DC voltage (720 V) argon plasma etching. After the etching process, we soaked the whole chip in 49\% hydrofluoric acid to remove the residue of alumina. Then the chip was deposited with another 5 nm Al$_2$O$_3$ to protect the Ag surface from oxidization in the later stages.

\subsubsection{Assembly with van der Waals heterostructures}

hBN flakes were mechanically exfoliated from the bulk crystals onto the silicon chip with a SiO$_2$ layer on top. Exfoliated MoSe$_2$ flakes were provided by the Quantum material press (QPress) facility in the Center for Functional Nanomaterials (CFN) at Brookhaven National Laboratory (BNL). The thickness of the hBN flakes and MoSe2 layer numbers are estimated based on the color contrast under optical microscopy. The 5 nm hBN (spacer)/ monolayer MoSe$_2$/ 10 nm hBN (for encapsulation) heterostructure was assembled in a transfer station built by Everbeing Int’l Corp. PDMS (Polydimethylsiloxane) and PC (Polycarbonate) was used to stamp and transfer all the flakes in a dry transfer method onto the fabricated device chip at 180$^{\circ}$C. 

\newpage

\subsection{\label{si:setup} 2. Experimental setup}

\begin{figure}[h]
\includegraphics[width=\columnwidth]{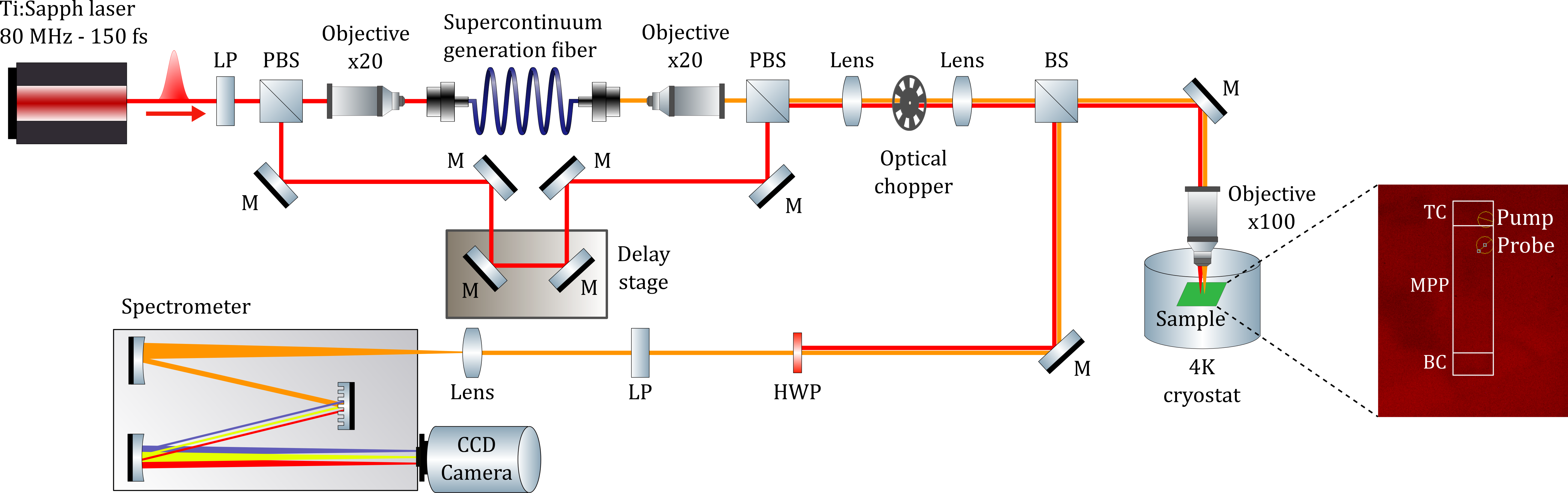}
\caption{\label{fig:setup} \textbf{Experimental setup.} Schematic of the experimental setup with a microscope image of the sample showing the relative positions of pump and probe beams. In the experimental setup: BS$=$Beam splitter, HWP$=$Half waveplate, LP$=$Linear polarizer, M$=$Mirror, PBS$=$Polarizing beam splitter. In the sample: BC$=$Bottom coupler, TC$=$Top coupler, WL$=$Waveguide lattice.}
\end{figure}

Figure~\ref{fig:setup} shows a schematic of our experimental setup.
The inset shows an image of the sample. For nonlocal AC Stark shift measurements, the pump is focused on the top coupler (TC). The TMD is probed $\sim3$ $\mu$m away from the pump. For local Stark shift measurements the pump and probe lie on the spot of the device. 

\newpage

\subsection{\label{si:wavelength} 3. Wavelength-dependent behavior of the MPP device}

\begin{figure}[h]
\includegraphics[width=\columnwidth]{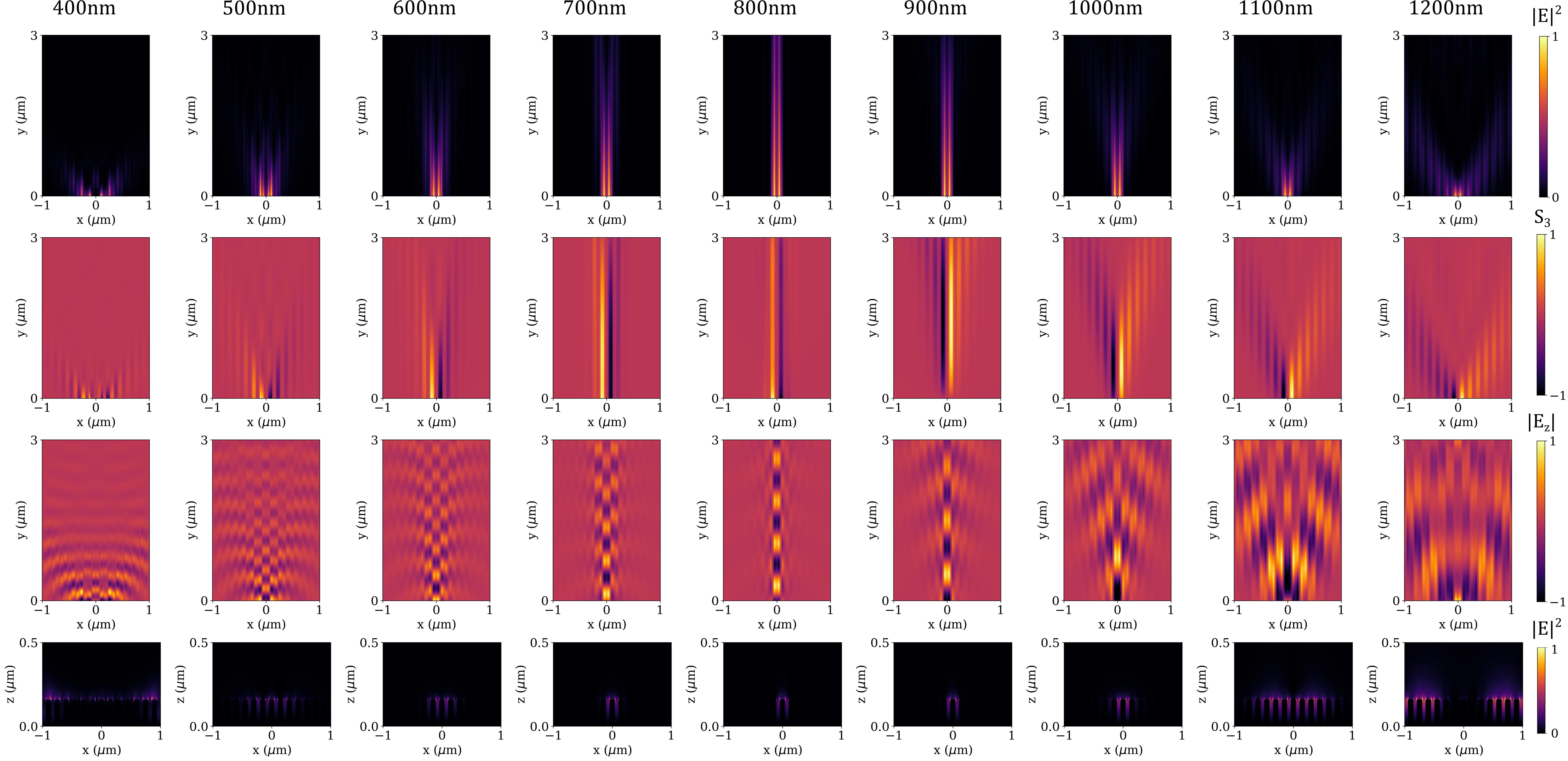}
\caption{\label{fig:wavelength} \textbf{Wavelength dependence of electric field intensity profile.} Row 1: Normalized $|E|^2$ monitored in the x-y-plane at 10 nm above the surface of the device for wavelengths between 400-1200 nm. We observe that the electric field does not diffract significantly between 700-800 nm. 
Row 2: Stokes parameter $S_3$ corresponding to panels in row 1. The MPP polarity switches signs at a transition wavelength between 700 nm and 800 nm. 
Row 3: Normalized $|E_z|$ corresponding to panels in rows 1 and 2. The dispersion transitions from hyperbolic to elliptical between 700 nm and 800 nm.
Row 4: Normalized $|E|^2$ at the x-z-plane cross-section of the device, positioned 1.5 $\mu$m away from the source.}
\end{figure}

We simulate the wavelength-dependent behavior of the MPP device using the FDTD method. We use a dipole aligned along the y-axis, located 10 nm above the surface to excite the MPP modes in the structure at in-plane coordinates $(x,y)=(0,-0.5)\mu$m. The period of the MPP device is $a=150$ nm, the height is $h=160$ nm, and the width is $w=90$ nm which corresponds to a duty cycle of 0.6.
Row 1 of Fig.~\ref{fig:wavelength} shows the normalized electric field intensity in the x-y plane monitored at 10 nm above the surface over a propagation length of 3 $\mu$m for wavelengths ranging from 400-1200 nm. We observe that the MPPs have negligible diffraction for the wavelength range 700-800 nm \cite{high2015visible}. 
Row 2 shows the Stokes parameter $S_3 = i(E_x E_z^* - E_z E_x^*)$ map corresponding to row 1, which shows us the degree of circular polarization of the MPPs as they propagate. We observe that the MPPs exhibit distinct and spatially resolved circularly polarized patterns. Moreover, the left- and right-diffracting modes flip circular polarization at a transition wavelength between 700-800 nm. This can be exploited to engineer valley-selective control of TMDs in such devices. 
At shorter wavelengths, the metasurface features a hyperbolic wavefront, while at longer wavelengths it becomes elliptical in nature as shown in row 3. This is a known and exciting topological transition that is typically used for hyperlensing and surfacing applications \cite{li2020collective}.
In row 4, we see the normalized electric field intensity at the x-z plane cross-section of the device, 1.5 $\mu$m away from the source for a similar range of wavelength. Once again, we observe that the MPPs have negligible diffraction for the wavelength range 700-800 nm.

\begin{figure}[h]
\centering
\includegraphics[width=0.9\columnwidth]{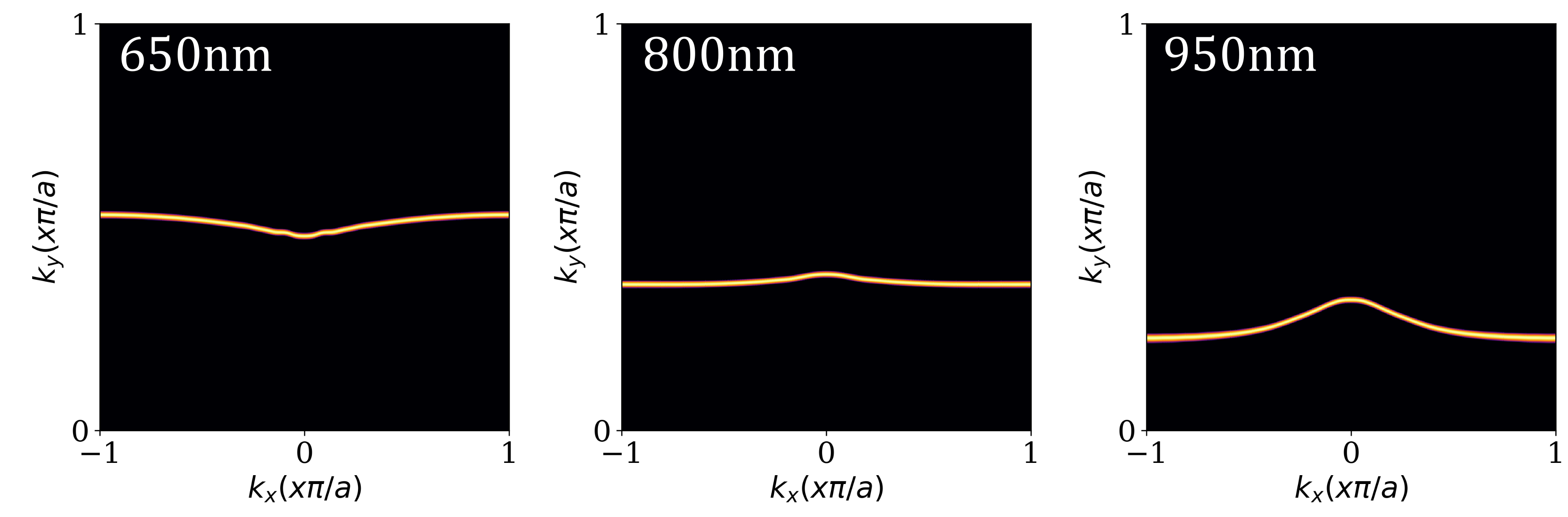}
\caption{\label{fig:IFC} \textbf{Iso-frequency contours of MPP.} The iso-frequency contours of the MPP at 650 nm, 800 nm, and 950 nm.}
\end{figure}

The transition from a hyperbolic to elliptical dispersion can also be observed in the iso-frequency contours, which are derived by calculating the Fourier transform of the simulated field profiles in Fig.~\ref{fig:wavelength}. Figure~\ref{fig:IFC} shows the iso-frequency contours at 650 nm, 800 nm, and 950 nm. We observe that $k_y$ is almost independent of $k_x$ for the wavelength of 800 nm where diffraction is negligible. At 650 nm, the dispersion is hyperbolic in nature, whereas at 950 nm it is elliptical.

\begin{figure}[h]
\centering
\includegraphics[width=0.55\columnwidth]{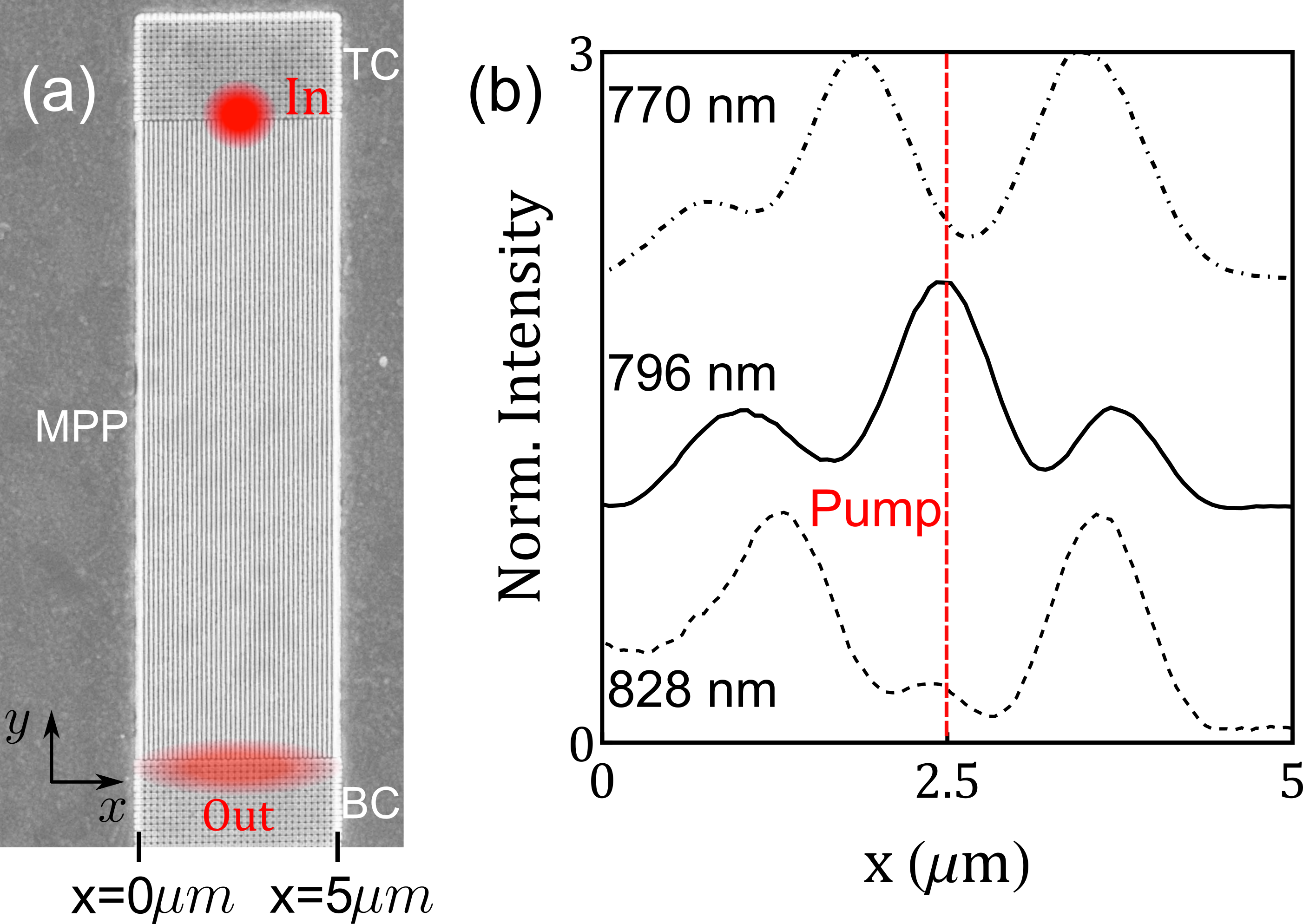}
\caption{\label{fig:grating_op} \textbf{Wavelength-dependent diffraction in MPP device.} (a) Measurement schematic where we shine light at the center of the TC, i.e., at $x=2.5$ $\mu$m, and spatially resolve the transmitted light at the BC. (b) The spatially resolved transmission from the BC for different wavelengths.}
\end{figure}

To experimentally characterize the wavelength-dependent diffraction of the MPP, we pump at the center of the top coupler (TC) along the x-direction, i.e., at $x=2.5$ $\mu$m as shown in Fig.~\ref{fig:grating_op}(a). We collect the diffracted beam from the bottom coupler (BC). Figure~\ref{fig:grating_op}(b) shows the spatially resolved transmission pattern observed at BC. For this device, we observe that the out-coupled light does not diffract significantly, and mostly comes out at $x=2.5$ $\mu$m around 796 nm, whereas the beam diffracts at higher and lower wavelengths, in agreement with previous studies \cite{high2015visible}.  

\newpage

\subsection{\label{si:IFC_duty_cycle} 4. Variation of dispersion with duty cycle}

\begin{figure}[h]
\centering
\includegraphics[width=\columnwidth]{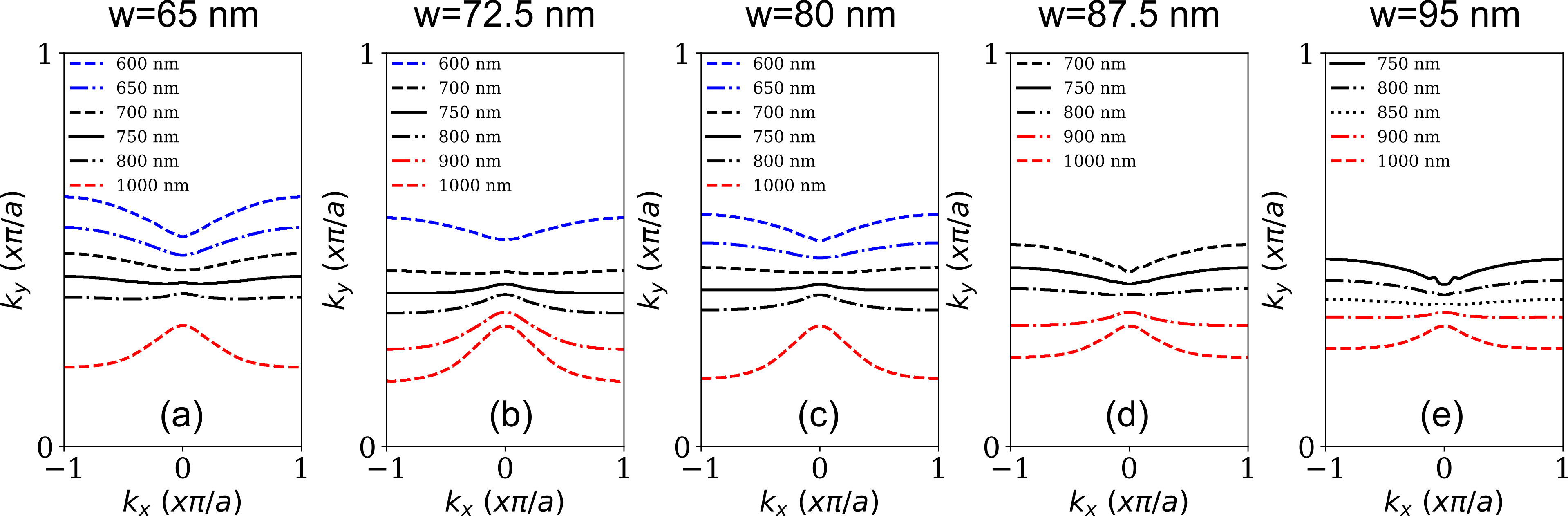}
\caption{\label{fig:IFC_DC} \textbf{Dependence of the dispersion on the duty cycle of MPP device.} The iso-frequency contours for a MPP with (a) $w=65$ nm, (b) $w=72.5$ nm, (c) $w=80$ nm, (d) $w=87.5$ nm, and (e) $w=95$ nm, respectively.}
\end{figure}

Due to fabrication imperfections, the duty cycle of the MPP device can deviate from the original design parameters. To study this effect, in Fig.~\ref{fig:IFC_DC} we show the change in the MPP dispersion with varying duty cycles. Using the FDTD method, we simulated the iso-frequency contours of the MPP device for (a) $w=65$ nm, (b) $w=72.5$ nm, (c) $w=80$ nm, (d) $w=87.5$ nm, and (e) $w=95$ nm. In all the simulations we chose period $a=150$ nm and height $h=160$ nm. One can observe that the wavelength at which the MPPs have minimal diffraction varies strongly with $w$. 

\newpage

\subsection{\label{si:theory} 5. Coupled mode theory}

\begin{figure}[h]
\includegraphics[width=0.7\columnwidth]{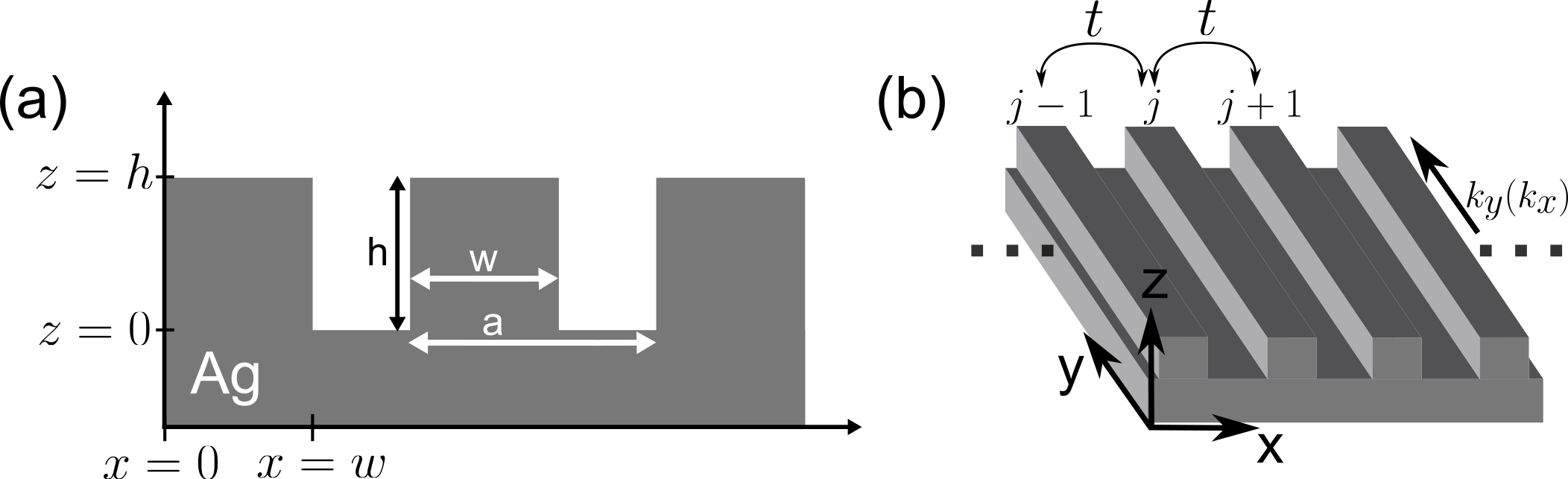}
\caption{\label{fig:theory1} \textbf{Coupled mode theory formalism.} (a) Cross-section of the metasurface device. (b) Coupled mode theory gives us $k_y(k_x)$ by assuming weak nearest-neighbor coupling between the grooves.}
\end{figure}

We follow Ref.~\cite{yariv2007photonics,pertsch2002anomalous,conforti2008subwavelength,locatelli2005diffraction, fan2006all} for this section.
The dispersion of the MPP can be modeled using coupled mode theory in the weak coupling limit. Figure \ref{fig:theory1} provides the schematic of the device. The dielectric function of the periodic metasurface can be written as $\epsilon(x,y,z) = \epsilon_{\rm{original}}(x,y,z) + \Delta \epsilon(x,y,z)$, where

\begin{equation}
    \Delta \epsilon(x,y,z) = \begin{cases}
                        0, &\text{for $z>h$} \\
                        \epsilon_0(\epsilon_{\rm{Air}} - \epsilon_{\rm{Ag}}) f(x), &\text{for $0<z<h$}\\

                        0, &\text{for $z<0$} 
                    \end{cases}
\end{equation}

\noindent is the perturbation on the original dielectric function $\epsilon_{\rm{original}}(x,y,z)$. $f(x)$ is a periodic square-wave modulation along the x-axis such that

\begin{equation}
    f(x) = \begin{cases}
                        0, &\text{for $0<z<w$} \\
                        1, &\text{for $w<z<a$}
                        
                    \end{cases}
\end{equation}

\noindent and $f(x)$ is Bloch-periodic, i.e., $f(x+a)=f(x)$. $f(x)$ can be expressed as a Fourier expansion as:

\begin{equation}
    f(x) = \sum_{l=-\infty}^{\infty} \phi_l e^{-i 2\pi lx/a},
\end{equation}

\noindent such that 

\begin{equation}
    \phi_l = \begin{cases}
                        \frac{1}{2}, &\text{for $l=0$} \\
                        0, &\text{for $l=$even} \\
                        \frac{i}{l\pi},
                        &\text{for $l=$odd}
                        
                    \end{cases}
\end{equation}

The electric field in this periodic structure can be expressed as a linear combination of the mode in each groove, i.e.,

\begin{equation}
    \Vec{E}(x,y,z,t) = \sum_j a_j(y) \Vec{E}_j(x-ja,z)e^{ik_jy},
\end{equation}

\noindent where $a_j(y)$ is the amplitude of the transverse electric field profile $\Vec{E}_j(x-ja,z)$ of the j$^{\rm{th}}$ groove, and $a$ is the periodicity in the x-direction. The amplitudes of the normal modes satisfy the relationship:

\begin{equation}
   i \dfrac{da_j}{dy} = \sum_l \sum_k t_{jk}^{(l)} a_k(y) e^{i(k_j - k_k -2\pi l/a)y},
\end{equation}

\noindent where $t_{jk}^{(l)}$ is the coupling constant (or the tunneling amplitude) between modes j and k originating from the l$^{\rm{th}}$ term in the Fourier expansion of the dielectric function. The exponential term in the differential equation is fast-oscillating and cancels out to zero unless there is resonant coupling between adjacent waveguides that satisfies the relationship $k_j - k_k -2\pi l/a = 0$. Thus, considering nearest neighbor coupling, we get

\begin{equation}
\label{eq:diff_eqn}
   i \dfrac{da_j}{dy} = t (a_{j-1}(y) + a_{j+1}(y)).
\end{equation}

\noindent The tunneling amplitude is given by $t=\kappa/P$ where \cite{yariv2007photonics,high2015visible}

\begin{align}
\kappa &= \dfrac{\omega}{4} \epsilon_0 (\epsilon_{\rm{Ag}}-\epsilon_{\rm{Air}}) \int \Vec{E}_j^* (x,z) \cdot \Vec{E}_k (x,z) \,dx \,dz,\\
P &= \dfrac{1}{4} \int (\Vec{E}_j^* (x,z) \times \Vec{H}_j (x,z) + \Vec{E}_j (x,z) \times \Vec{H}_j^* (x,z) ) \cdot \hat{y} \,dx \,dz.
\end{align}

\begin{figure}[b]
\includegraphics[width=0.55\columnwidth]{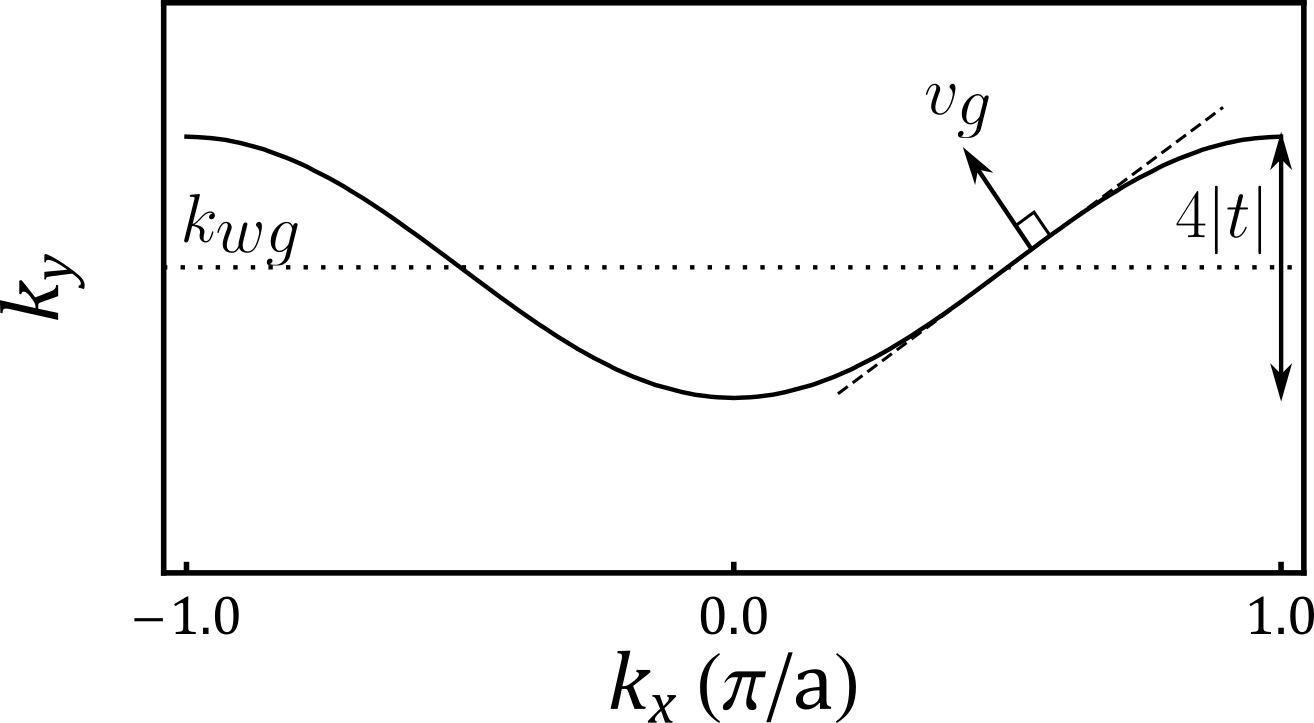}
\caption{\label{fig:theory2} \textbf{Iso-frequency contour.} Hyperbolic dispersion relation of the metasurface.}
\end{figure}

The solution to Eqn. \ref{eq:diff_eqn} is given by \cite{pertsch2002anomalous,conforti2008subwavelength}:

\begin{equation}
    k_y(k_x) = k_{\rm{wg}} + 2t \cos(k_x a),
\end{equation}

\noindent where $k_{\rm{wg}}$ is the propagation constant of the waveguide. 
Depending on the wavelength of light and duty cycle of the metasurface, one can have either positive or negative coupling constant $t$ \cite{locatelli2005diffraction, fan2006all}. The negative coupling constant $t$ gives the hyperbolic dispersion relationship as shown in Fig.~\ref{fig:theory2}. This qualitatively matches with the iso-frequency contours simulated using the FDTD method. The group velocity $v_g = \frac{d\omega}{dk}$ lies orthogonal to the iso-frequency contour, and the slope gives the angle of refraction $\theta = \tan^{-1}(\frac{dk_y}{dk_x})$.

The 1D lattice along the x-direction gives rise to the tight-binding expression $\sim \cos(k_x a)$. It is important to note that the momentum along the x- and y-directions are not independent, but phase-locked among adjacent grooves, making the iso-frequency contour relationship resemble a dispersion relationship for the MPP.

\newpage

\subsection{\label{si:budget} 6.  Power efficiency calibration and field confinement}

\subsubsection{A. Power efficiency calculations}

\begin{figure}[h]
\includegraphics[width=\columnwidth]{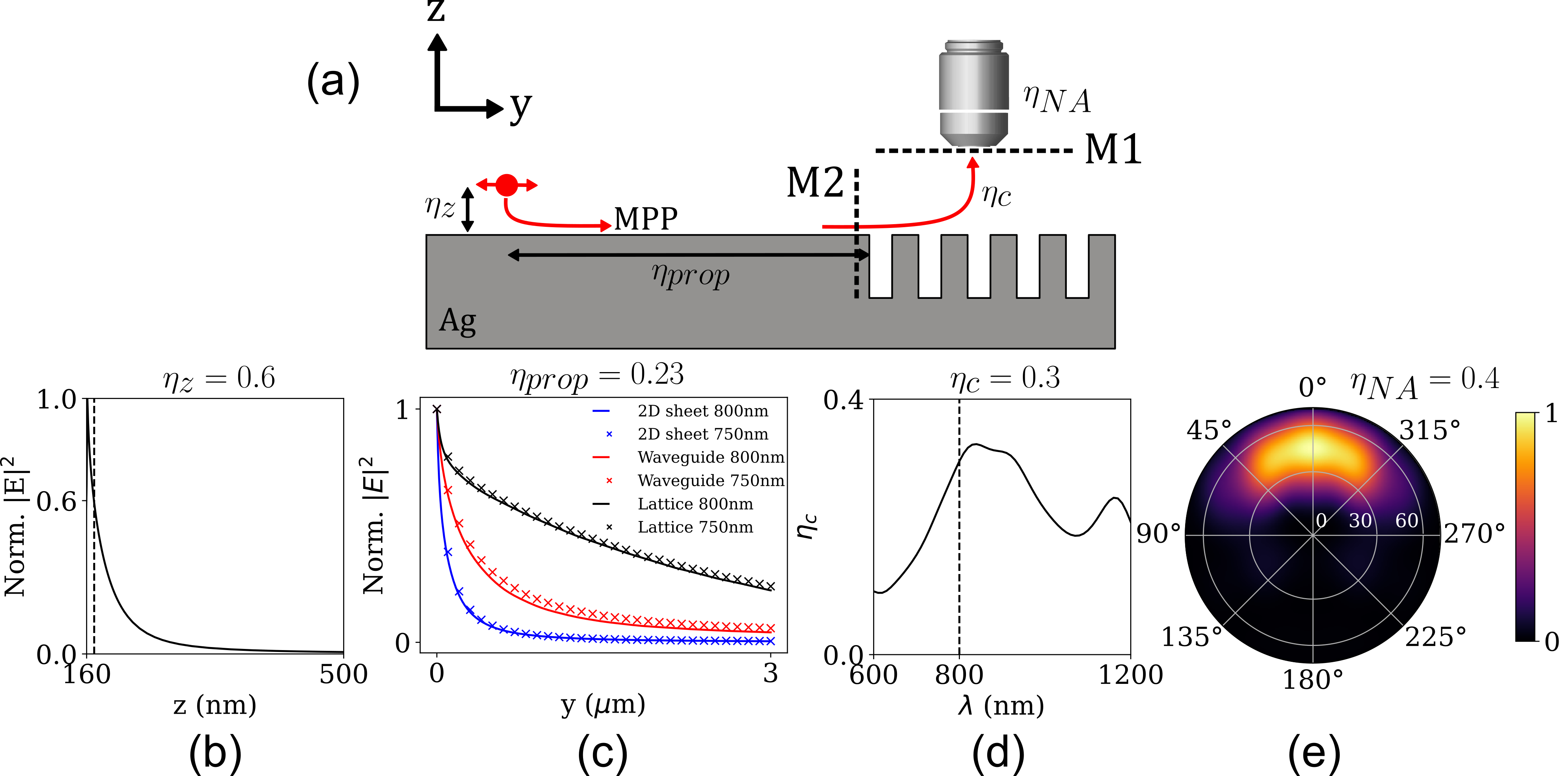}
\caption{\label{fig:power budget} \textbf{Power efficiency calculations.} The simulation scheme for estimating the power efficiency of the nonlocal AC Stark shift scheme. (a) The different losses captured by the efficiencies $\eta_z$, $\eta_{prop}$, $\eta_c$, and $\eta_{NA}$ at 800 nm. (b) Decay of $|E|^2$ at 800 nm from the surface of the device at $z=160$ nm. The dashed line shows a cut at a distance of 10 nm above the surface, where $\eta_z$ = 0.6. (c) Decay of $|E|^2$ as the MPPs propagate in the metasurface lattice (black), in comparison to propagation in a single waveguide of similar dimension (red), or an unpatterned 2D sheet of silver (blue). (d) The wavelength-dependent coupling efficiency of the MPPs to far-field. (e) Far-field emission pattern from monitor M1 in panel (a).}
\end{figure}

We simulate the power delivery efficiency to the TMD monolayer (ML) through our MPP device by simulating the time-reversal counterpart using the FDTD method as shown in Fig.~\ref{fig:power budget}(a). We excite the MPP modes in the metasurface using a dipole 10 nm above the surface with efficiency $\beta$. The emission wavelength of the dipole is 800 nm.

The exponential decay of $|E|^2$ as a function of vertical distance from the surface of the MPP device is captured in parameter $\eta_z$ as shown in Fig.~\ref{fig:power budget}(b). The surface of the MPP device is at $z=160$ nm. The TMD is located at a distance of 10 nm above the surface because of the presence of Al$_2$O$_3$ to prevent the silver from oxidizing and hBN encapsulation, which yields an $\eta_z=0.6$, as marked by the dashed vertical line. Reducing the thickness of Al$_2$O$_3$ and hBN can enhance the efficiency.

The MPP modes then propagate along the lattice for 3 $\mu$m. The propagation efficiency is given by $\eta_{prop}$. Figure~\ref{fig:power budget}(c) shows the decay of $|E|^2$ as the MPPs propagate in the metasurface lattice (black), in comparison to propagation in a single waveguide of similar dimension (red), or an un-patterned 2D sheet of silver (blue). From this, we estimate a propagation efficiency of $\eta_{prop}=0.23$. 

Next, the MPPs are coupled to the far-field through the cylindrical couplers. We calculate the coupling efficiency by monitoring the electric field intensity transmitted through monitor M1 (placed 600 nm above the coupler) and monitor M2 (placed immediately before the MPPs reach the coupler). Figure~\ref{fig:power budget}(d) shows the wavelength-dependent coupling efficiency $\eta_c$. For 800 nm, we calculate $\eta_c=0.3$. 

Figure~\ref{fig:power budget}(e) shows the normalized electric field intensity in the far-field for 800 nm. Assuming a vertically placed objective of numerical aperture 0.7 (the one used in our experiment), we calculate $\eta_{NA}$ as the fraction of power inside the numeric aperture. This gives us $\eta_{NA}=0.4$.

Considering the different loss mechanisms, we get a total efficiency of $\eta = \beta \times \eta_z \times \eta_{prop} \times \eta_c \times \eta_{NA} = 0.33\%$.

\newpage

\subsubsection{B. Electric field confinement}

\begin{figure}[h]
\includegraphics[width=0.7\columnwidth]{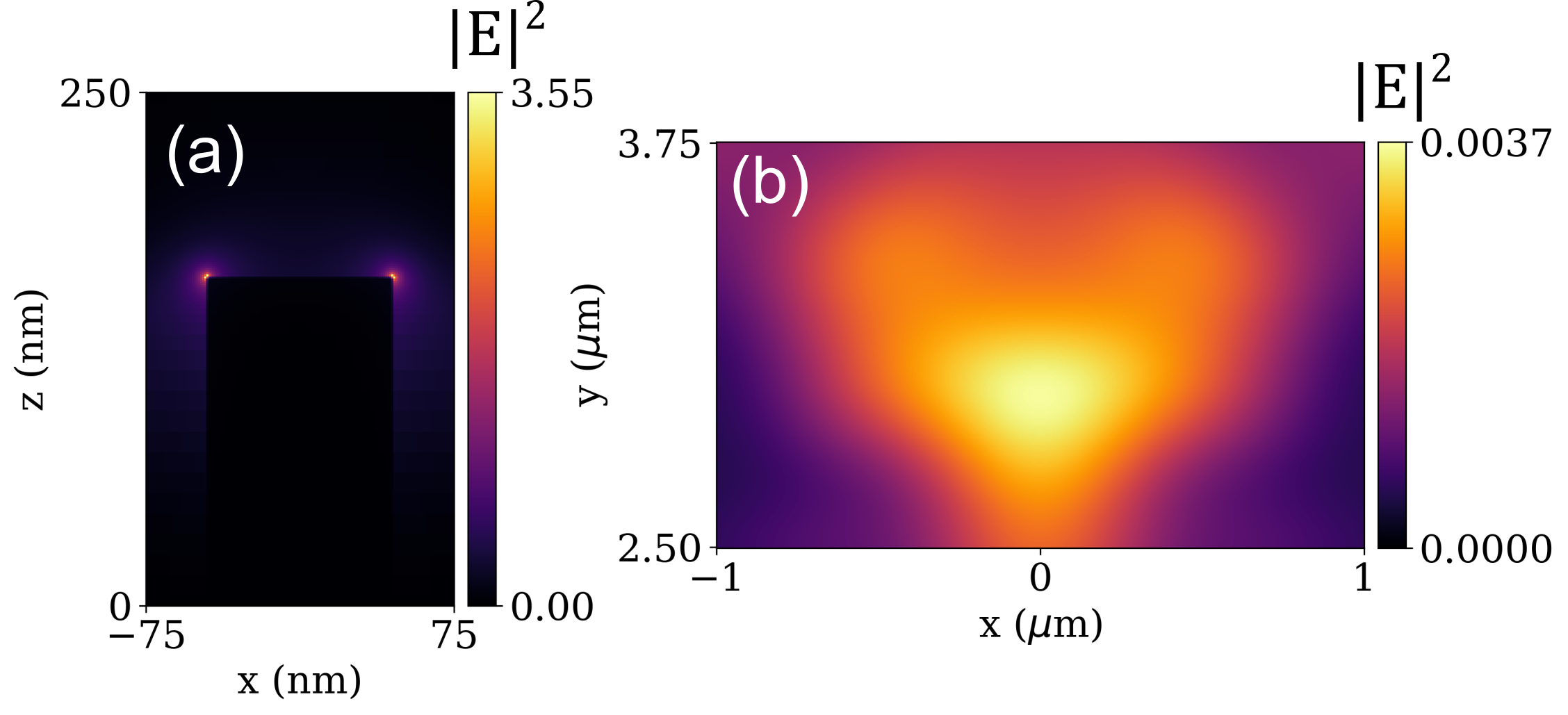}
\caption{\label{fig:confinement} \textbf{Electric field intensity confinement.} (a) Electric field intensity $|E|^2$ at monitor M2 in Fig.~\ref{fig:power budget}. (b) Electric field intensity $|E|^2$ at monitor M1 in Fig.~\ref{fig:power budget}.}
\end{figure}

To estimate the electric field confinement provided by the MPP device, we monitor the electric field intensity at M1 and M2, as shown in Fig.~\ref{fig:confinement}. This gives us a maximum electric field intensity confinement by a factor of $\sim960$. However, taking into account the propagation loss and decay of the field intensity away from the surface, an exciton located 3 $\mu$m away from the coupler is expected to experience a field intensity enhancement by a factor of $960 \times \eta_{prop} \times \eta_z \approx 130$. However, this is just an upper bound on the electric field confinement calculated for an ideally placed dipole with a fixed polarization.

\newpage

\subsubsection{C. Sensitivity to dipole position and polarization}

\begin{figure}[h]
\includegraphics[width=\columnwidth]{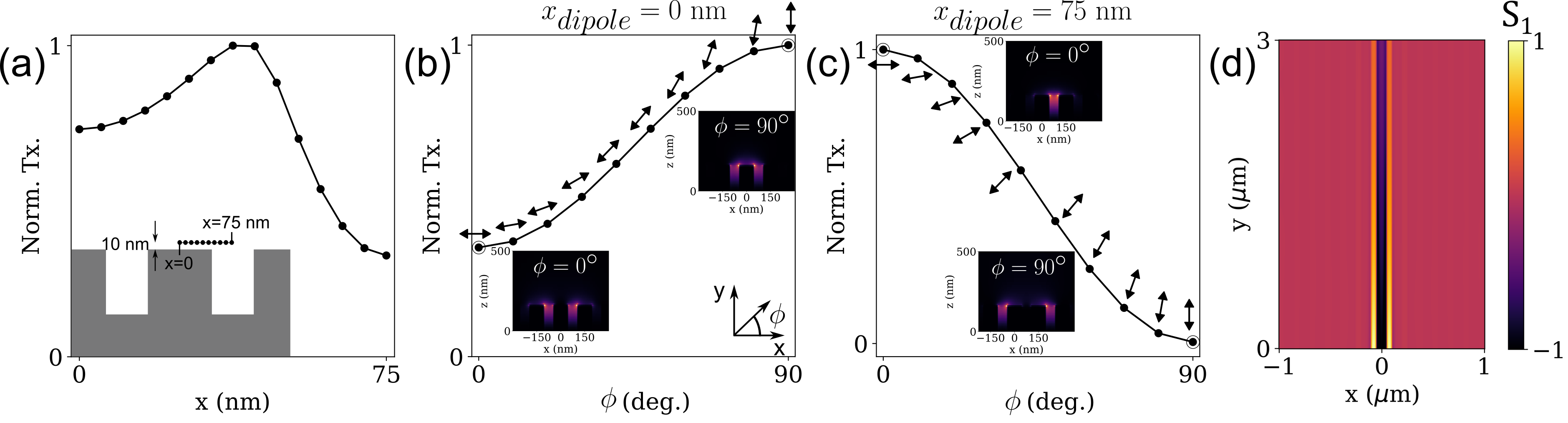}
\caption{\label{fig:pos_pol} \textbf{Sensitivity to dipole position and polarization.} (a) Normalized power transmission through the MPP device as a function of dipole position. (b-c) Normalized power transmission through the MPP device as a function of dipole polarization for dipole positions of $x_{dipole}=0$ nm, 75nm respectively. (d) Stokes parameter $S_1$ in the x-y plane at 10 nm above the MPP device surface. For these simulations, we use a dipole with an emission wavelength of $\lambda=800$ nm.}
\end{figure}

For the FDTD simulations, we have used an in-plane dipole aligned along the y-axis ($\phi=90^{\circ}$) to excite the MPP modes. However, we note that the dipoles of the TMD monolayer are located throughout the x-y plane in arbitrary orientations. To estimate the effect of dipole position and orientation, we perform additional FDTD simulations.
Figure~\ref{fig:pos_pol}(a) shows the normalized power transmitted through the MPP device as a function of dipole location along the y-axis. We note that this closely resembles the normalized electric field intensity at the location of the TMD ML as shown by the red curve in Fig.~\ref{schematic}(b). The inset shows the positions of the dipoles used for the sweep.
Figure~\ref{fig:pos_pol}(b-c) shows the normalized power transmitted through the MPP device as a function of dipole orientation $\phi$ for a dipole located at $x_{dipole}=0$ nm, 75nm respectively. The insets show the normalized electric field intensity $|E|^2$ in the x-z cross-section of the device for different polarizations of the dipole. 
Figure~\ref{fig:pos_pol}(d) shows the normalized Stokes parameter $S_1=|E_x|^2 - |E_z|^2$ above the MPP device for $x_{dipole} = 0$ nm and $\phi = 90^{\circ}$. The electric field is predominantly z-polarized over the silver waveguides, whereas it is x-polarized in the air gaps. The sensitivity of the electric field to different positions and polarizations of dipoles is expected to give a smaller AC Stark shift than what would be expected from the upper bound of the electric field confinement as calculated earlier.

\newpage

\subsection{\label{si:multiple_dipoles} 7. Exciting multiple waveguides in the metasurface}

\begin{figure}[h]
\includegraphics[width=0.85\columnwidth]{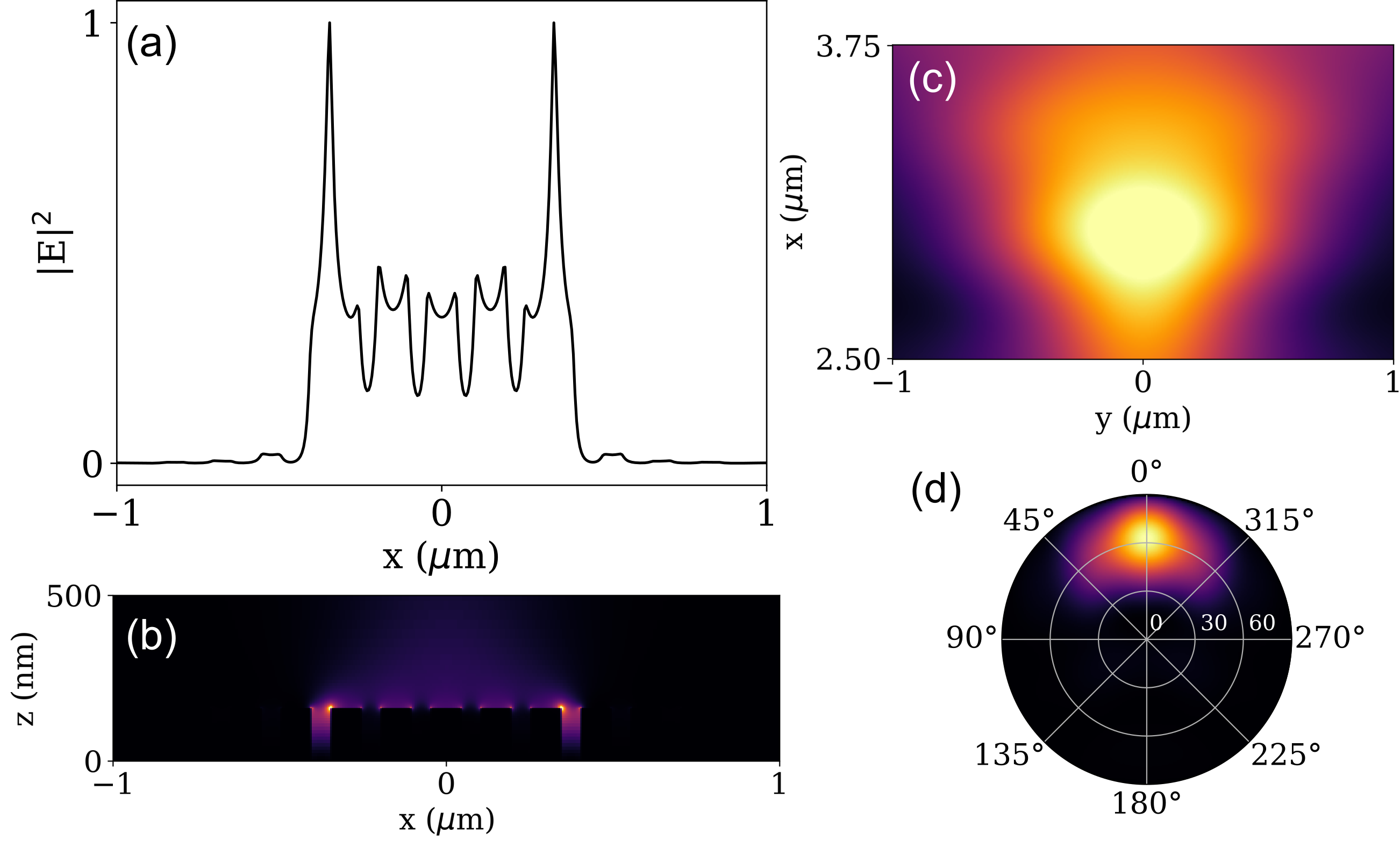}
\caption{\label{fig:multiple_dipoles} \textbf{Diffraction-limited excitation of multiple grooves.} (a) Normalized electric field intensity $|E|^2$ 10 nm above the surface if MPPs are excited in 5 adjacent grooves. This is used to periodically modulate the potential landscape imprinted on the TMD ML at a sub-wavelength scale. (b) Normalized $|E|^2$ at the cross-section of the device. (c) Normalized $|E|^2$ monitored in the x-y plane 600 nm above the gratings. (d) The far-field radiation pattern obtained from the intensity profile in panel (c). For all the panels the wavelength of MPPs is 800 nm. }
\end{figure}

Figure~\ref{fig:multiple_dipoles} shows the FDTD simulations for the excitation of MPPs in multiple grooves simultaneously in our diffraction-limited excitation scheme. Here we have excited the MPPs in 5 adjacent grooves using 5 identical dipoles. In Fig.~\ref{fig:multiple_dipoles}(a) we can observe that the TMD experiences a sub-wavelength periodic modulation of electric field intensity 10 nm above the surface of the MPP device. Figure~\ref{fig:multiple_dipoles}(b) shows the normalized $|E|^2$ at the x-z cross-section of the PM device. However, we note that in an ideal scenario, the pump beam has a Gaussian spatial profile and that would modify the exact spatial intensity profile. Figure~\ref{fig:multiple_dipoles}(c) shows the normalized $|E|^2$ monitored in the x-y plane 600 nm above the gratings, and Fig.~\ref{fig:power budget}(d) shows the corresponding far-field radiation pattern. Upon comparisons to their counterparts for a single groove excitation as shown in Fig.~\ref{fig:power budget}(e) and Fig.~\ref{fig:confinement}(b), we note that exciting multiple grooves do not modify the far-field significantly.

\newpage

\subsection{\label{si:defect} 8. Exciting a single waveguide in the metasurface}

\begin{figure}[h]
\includegraphics[width=\columnwidth]{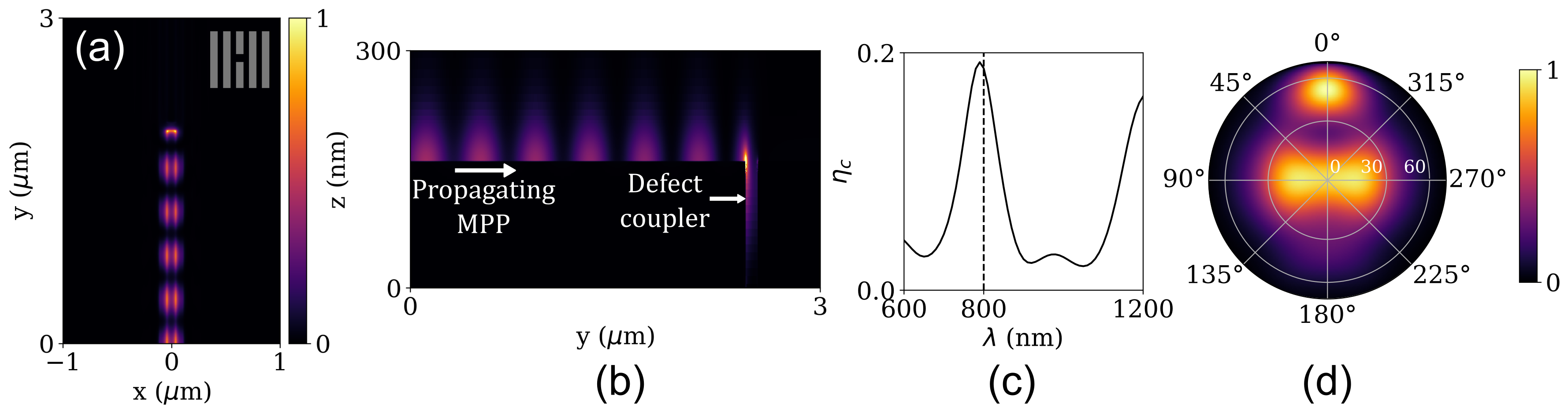}
\caption{\label{fig:defect} \textbf{Sub-diffraction excitation of a single groove using a defect.} (a) Electric field intensity 10 nm above the surface of a device with a single point-defect coupler.
(b) A side-view showing $|E|^2$ of the propagating MPPs being scattered out of the defect coupler. (c) The wavelength-dependent output coupling efficiency. (d) The far-field emission pattern of the defect coupler.}
\end{figure}

The MPP platform also allows us to excite a single groove using a defect coupler, as demonstrated in Ref.~\cite{high2015visible}. This enables one to address a mesoscopic number of excitons on the TMD in a $\sim100$ nm scale. To consider the viability of such an excitation scheme, we perform FDTD simulations to estimate the power efficiency as shown in Fig.~\ref{fig:defect}. We perform the reciprocal simulation by exciting the MPPs with a dipole and monitoring the scattered light out of the defect. Figure~\ref{fig:defect}(a) shows the electric field intensity 10 nm above the surface of a device with a single point-defect coupler. The 90 nm $\times$ 90 nm defect is located at $x=2$ nm. The inset shows a simplified schematic of the simulated device. Figure~\ref{fig:defect}(b) shows the propagating MPPs being scattered off the defect coupler from the side view cross-section. 
In this scheme, $\eta_z$ and $\eta_{prop}$ will remain unchanged with respect to our main scheme discussed in the previous sections. The coupling efficiency $\eta_c$ is calculated to be 19\%, and $\eta_{total}$ is 0.25\% (considering NA of 0.7) as shown in Fig.~\ref{fig:defect}. 

\begin{figure}[h]
\includegraphics[width=\columnwidth]{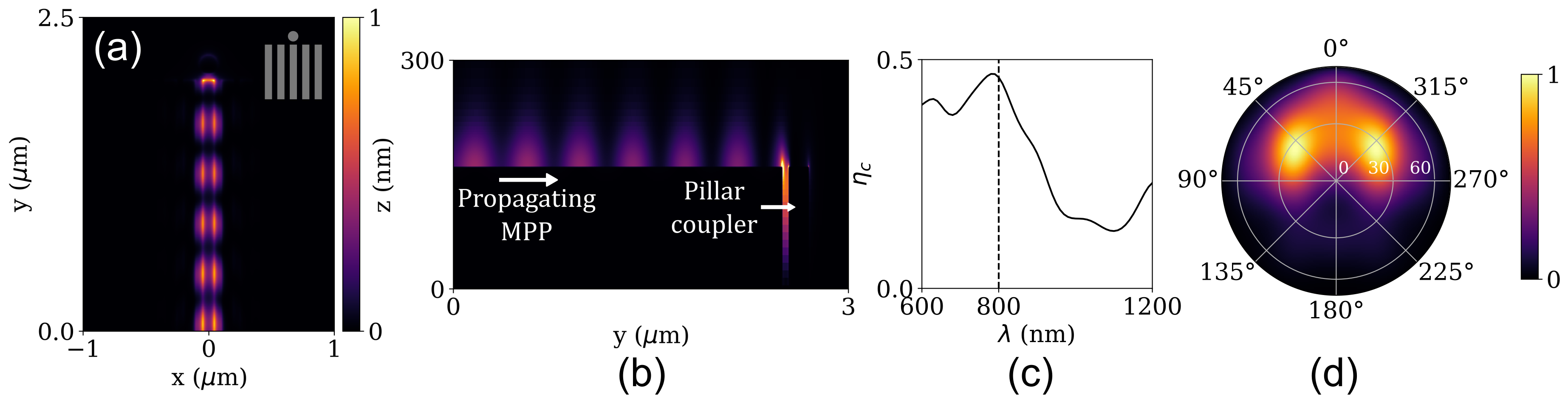}
\caption{\label{fig:pillar} \textbf{Sub-diffraction excitation of a single groove using a nano-pillar.} (a) Electric field intensity 10 nm above the surface of a device with a single nano-pillar.
(b) A side-view showing $|E|^2$ of the propagating MPPs being scattered out of the single nano-pillar. (c) The wavelength-dependent output coupling efficiency. (d) The far-field emission pattern of the single nano-pillar.}
\end{figure}

Another similar method could be using a single nano-pillar as shown in Fig.~\ref{fig:pillar}. Using similar simulations as in the case of the defect couplers we estimate $\eta_c=46\%$ and $\eta_{total}=0.7\%$. 

\newpage

\begin{figure}[h]
\includegraphics[width=0.7\columnwidth]{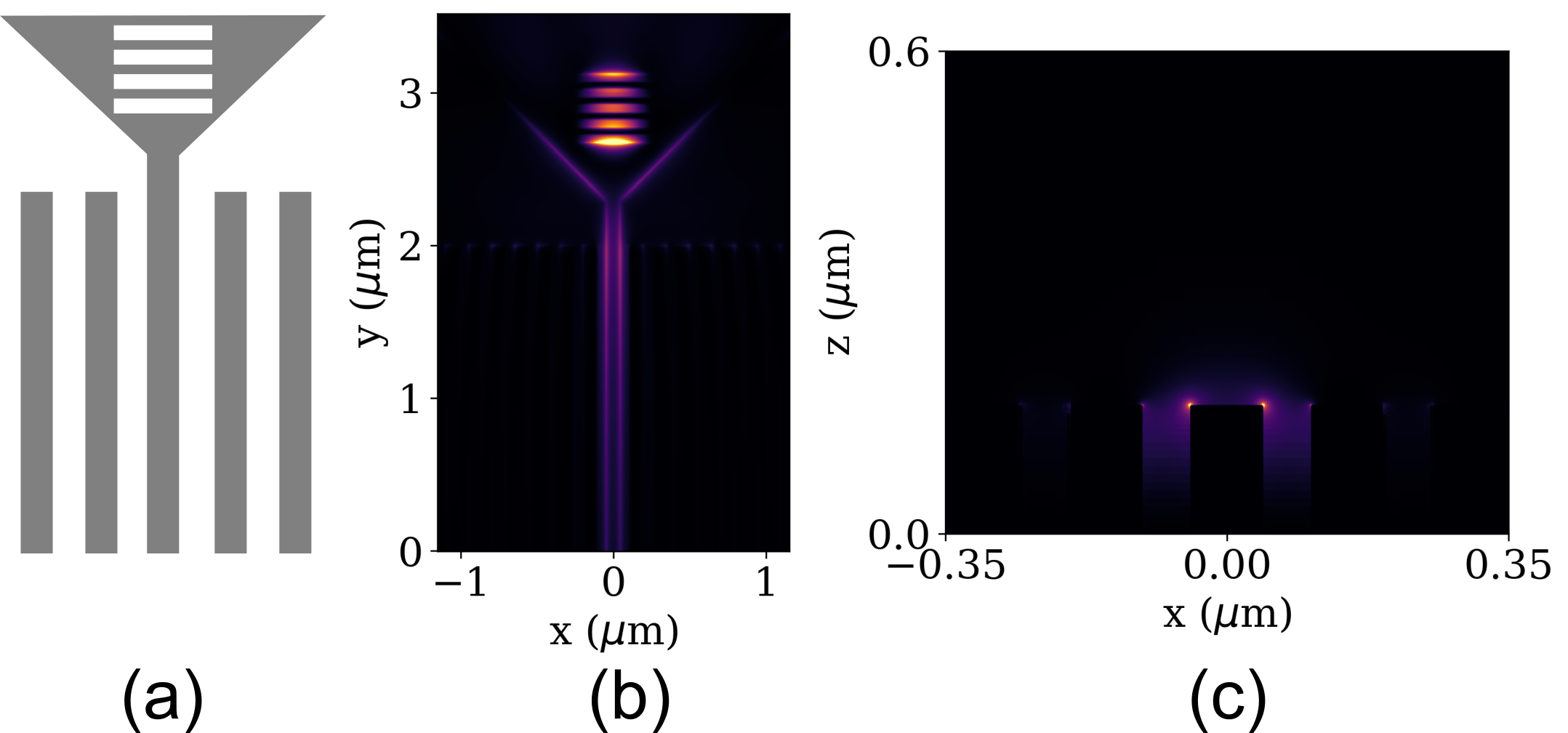}
\caption{\label{fig:taper_grating} \textbf{Sub-diffraction excitation of a single groove using a tapered grating.} (a) A schematic of the device to excite a single groove. (b) The electric field intensity profile at 10 nm above the surface of the device.
(c) Electric field intensity at the cross-section of the MPP device showing electric field tightly confined to a single groove at 800 nm.}
\end{figure}

Another method to couple light into a single groove is using grating patterns inspired by the Ref.~\cite{cheng2015topologically}. Figure~\ref{fig:taper_grating} shows a simulation of this scheme. In the simulation, a Gaussian beam is incident on the grating couplers which then excites the MPPs in a single groove. Figure~\ref{fig:taper_grating}(a) shows a simplified schematic of the device. Figure~\ref{fig:taper_grating}(b-c) shows the electric field intensity $|E|^2$ for 800 nm at a plane 10 nm above the surface of the device, and at the cross-section of the MPP device, respectively. We perform additional simulations to excite the MPPs in a single groove with a dipole, and for this device architecture we estimate $\eta_c=30\%$ and $\eta_{total}=0.1\%$.

\newpage


\subsection{\label{si:control} 9. Control experiments}

\subsubsection{A. Nonlocal AC Stark shift on MPP device and on 2D sheet}

\begin{figure}[h]
\includegraphics[width=0.7\columnwidth]{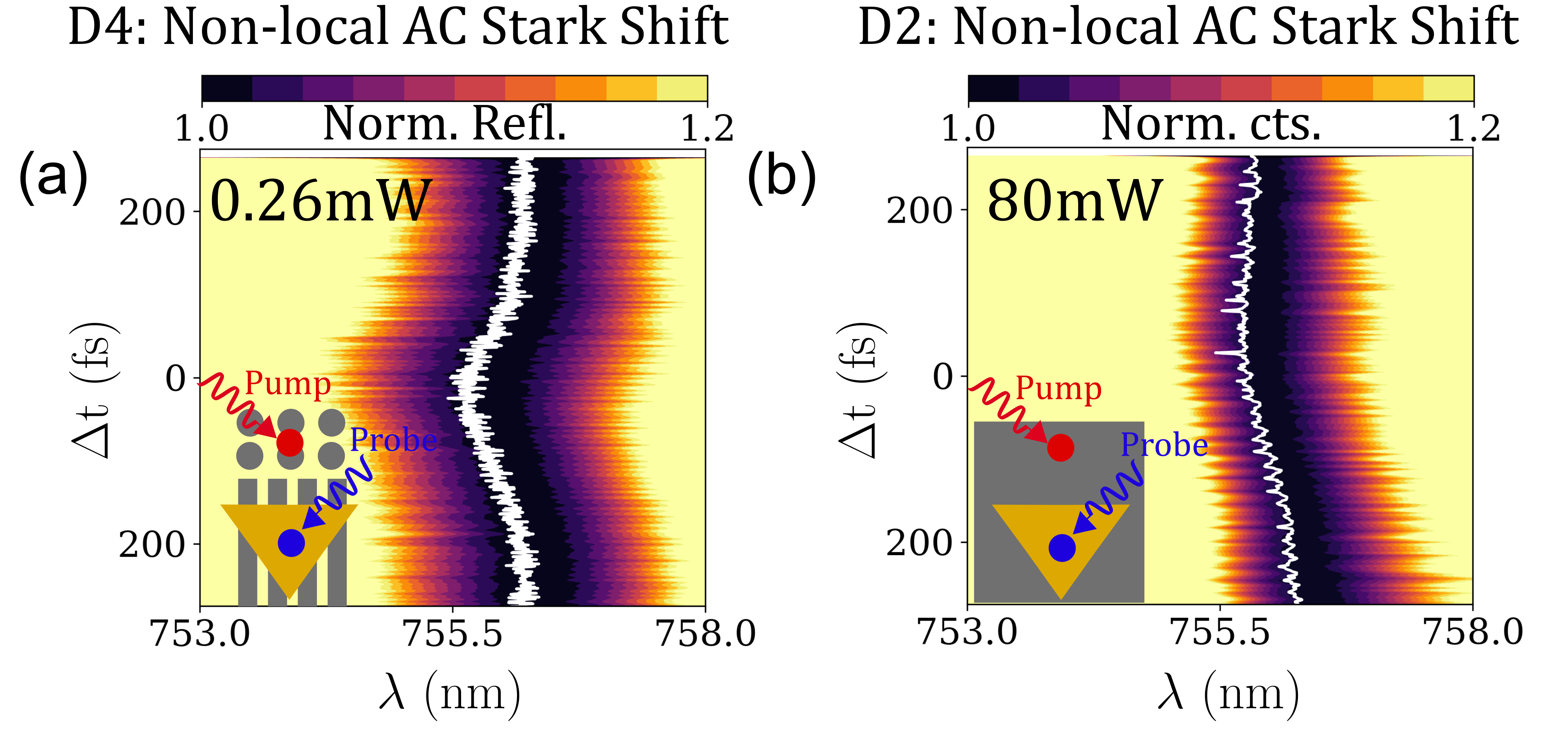}
\caption{\label{fig:control_exp} \textbf{Nonlocal AC Stark shift : control experiments.} (a-b) Nonlocal AC Stark shift measured in devices D4 and D2, respectively. The insets show schematics of the measurement schemes.}
\end{figure}

To confirm that the observed nonlocal AC stark shift is exclusively due to MPPs instead of the presence of the tail of a very strong pump laser at the position of the probe, we perform the following control experiments. We start with devices D4 and D2 on the same chip. While D4 has a MoSe$_2$ monolayer stacked on the MPP device, D2 has a MoSe$_2$ monolayer stacked on an unpatterned silver sheet. Using the same pump power of 80 mW (i.e. a pump power of 260 $\mu$W at the probing spot for the MPP device), and a pump-probe spatial separation of $\sim$3 $\mu$m, we perform the nonlocal AC stark shift measurement on both samples by scanning over the time delay of the pump and probe laser pulses as shown in Fig.~\ref{fig:control_exp}(a-b). Importantly, we only observe AC Stark shift in device D4. The absence of AC Stark shift in device D2 is likely due to weaker field confinement and smaller $\eta_{prop}$ in 2D silver sheets. Moreover, in device D2, we notice a redshift for negative values of $\Delta t$, very likely originating from the heating of the sample by the strong pump laser.

\begin{figure}[h]
\includegraphics[width=0.3\columnwidth]{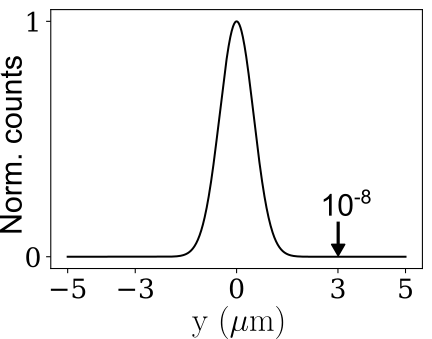}
\caption{\label{fig:pump-gauss} \textbf{Spatial profile of pump.} The normalized spatial profile of Gaussian pump with an FWHM of 1 $\mu$m.}
\end{figure}

Figure~\ref{fig:pump-gauss} shows that a Gaussian pump with an FWHM of 1 $\mu$m has 8 orders of magnitude less intensity 3 $\mu$m away from its peak.

\newpage

\subsubsection{B. Local AC Stark shift on MPP device}

\begin{figure}[h]
\includegraphics[width=0.75\columnwidth]{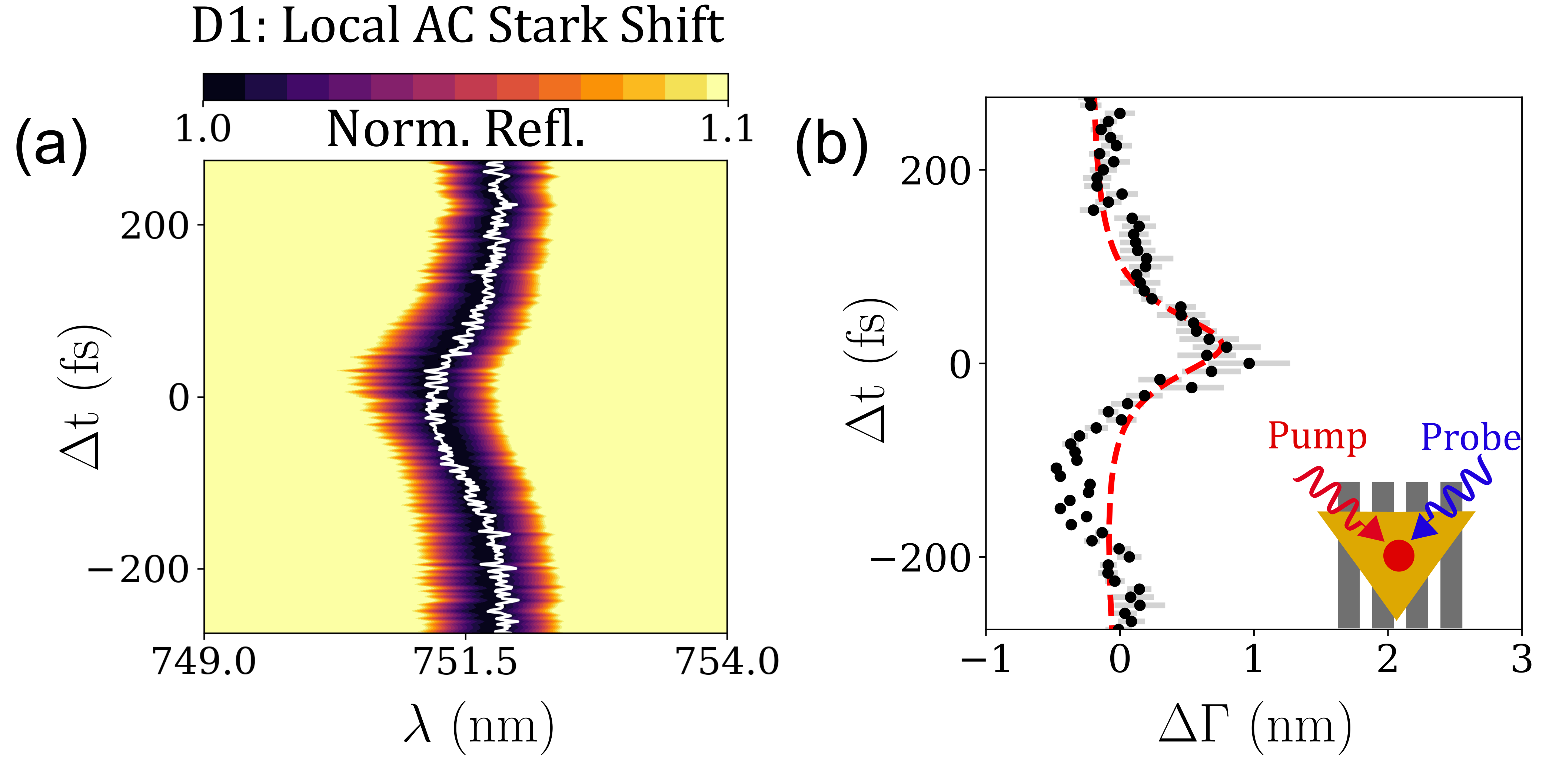}
\caption{\label{fig:Local_PMS} \textbf{Local AC Stark shift on MPP device.} (a) Local AC Stark shift measured in device D1. (b) Change in linewidth extracted by fitting the spectrum in panel (a). The inset shows a schematic of the measurement scheme.}
\end{figure}

To confirm the observation of linewidth broadening in the nonlocal AC Stark shift measurement as a signature of the sub-diffraction modulation of the electric field intensity on the TMD, we also perform local AC Stark shift measurements on the MPP device as shown in Fig.~\ref{fig:Local_PMS}. Notably, in this case, the linewidth of the excitons significantly broadens by $\sim1$ nm at $\Delta t=0$ alongside a Stark shift of $\sim0.65$ nm. This is in contrast to the absence of linewidth broadening observed in local AC Stark shift measurement. This strongly suggests that the linewidth broadening is a signature of a sub-wavelength modulation of the electric field intensity on the ML due to the presence of the lattice. The excitons that sense a strong electric field intensity have a large AC Stark shift compared to other excitons, leading to the broadening of the linewidth.

\newpage

\subsubsection{C. Local AC Stark shift on silica substrate}

\begin{figure}[h]
\includegraphics[width=0.75\columnwidth]{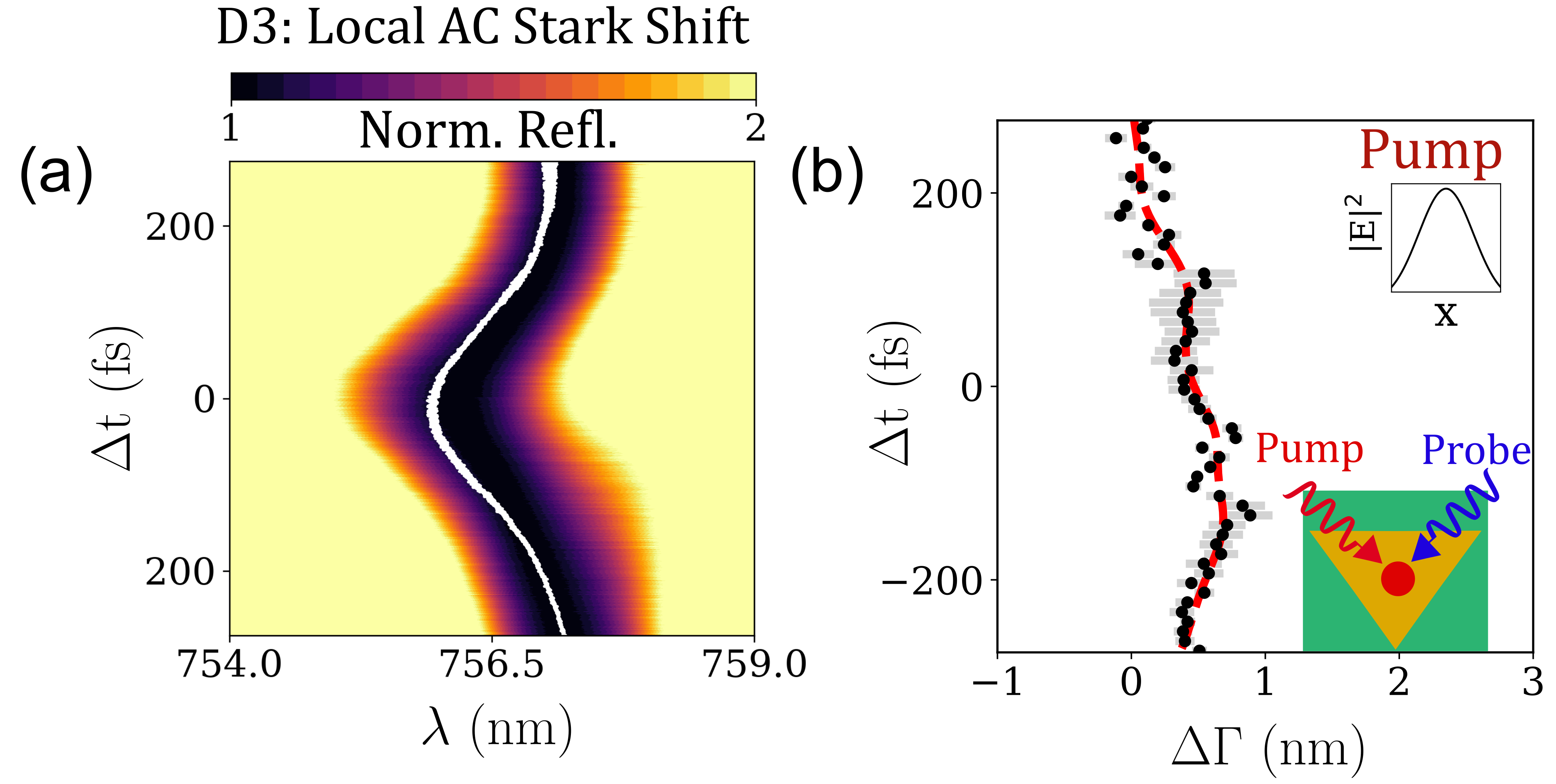}
\caption{\label{fig:Local_SiO2} \textbf{Local AC Stark shift on SiO$_2$.} (a) Local AC Stark shift measured in device D3. (b) Change in linewidth extracted by fitting the spectrum in panel (a). The inset shows a schematic of the measurement scheme and the diffraction-limited optical pump.}
\end{figure}

We measured local AC Stark shift on a TMD ML placed on an unpatterned silica substrate. From Fig.~\ref{fig:Local_SiO2}(a) we observe an AC Stark shift of $\sim1.1$ nm for a pump power of 60 mW. Importantly, we do not observe any strong modulation of linewidth at $\Delta t = 0$ similar to our observation in local AC Stark shift measurement performed on unpatterned silver.

\newpage

\subsubsection{D. Pump power and detuning dependence of AC Stark shift}

\begin{figure}[h]
\includegraphics[width=0.7\columnwidth]{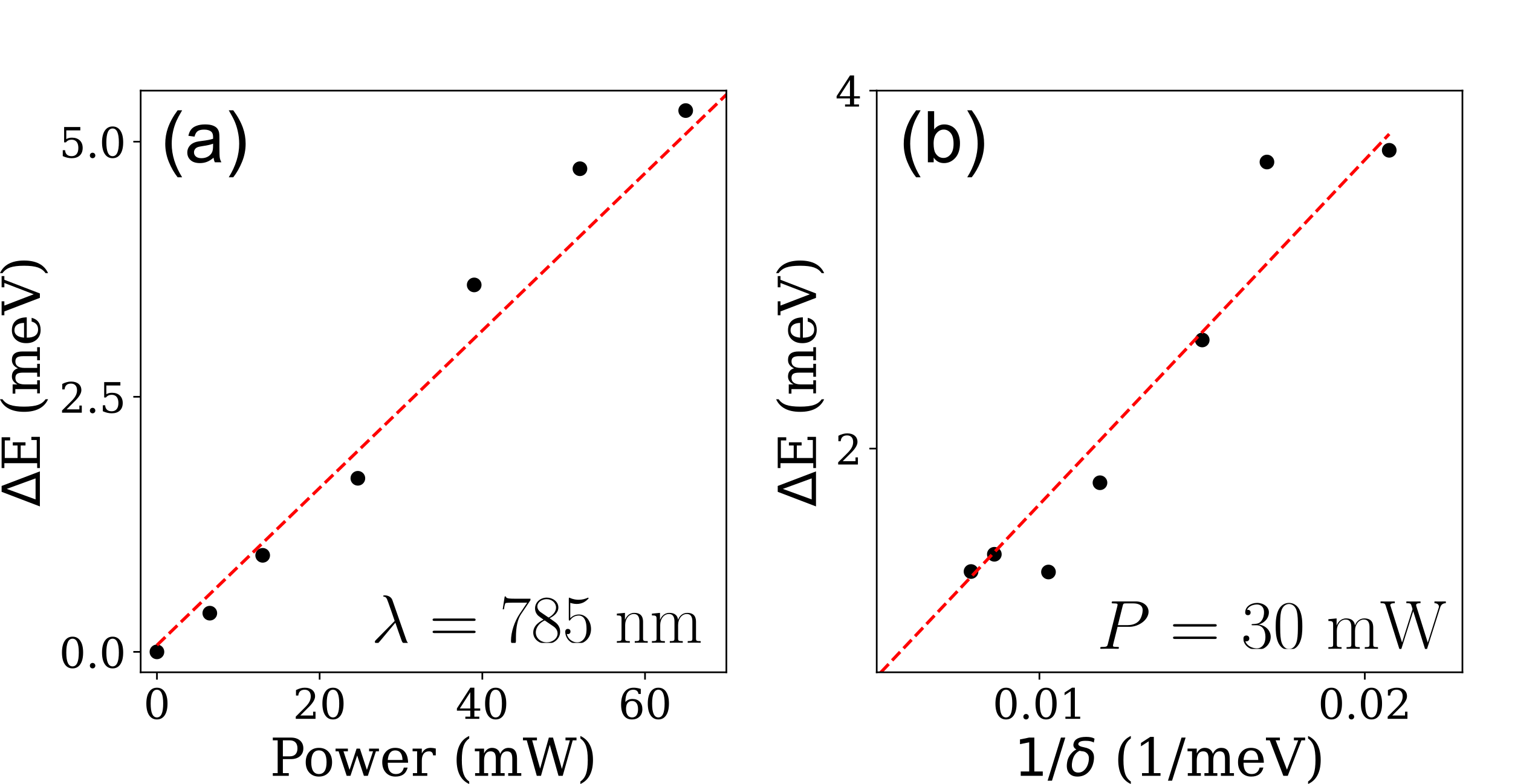}
\caption{\label{fig:power_detuning} \textbf{Pump power and detuning dependence of AC Stark shift.} (a) AC Stark shift $\Delta E$ as a function of pump power for a fixed pump wavelength. (b) AC Stark shift $\Delta E$ as a function of reciprocal pump detuning $1/\delta$ for a fixed pump power.}
\end{figure}

To confirm the expected linear and inverse dependence of the AC Stark shift on pump power and detuning, respectively \cite{sie2015valley,uto2024interaction}, we measured the AC Stark shift using a local pump-probe scheme as a function of pump power and detuning. As expected, we find that the AC Stark shift $\Delta E$ scales linearly with pump power for a fixed pump wavelength as shown in Fig.~\ref{fig:power_detuning}(a), since $\Delta E \sim \Omega_{\rm{pump}}^2$, where $\Omega_{\rm{pump}}$ is the Rabi frequency of the pump. Next, we sweep over the wavelength of the pump laser while keeping its power constant to study the dependence of the AC Stark shift on pump detuning. Figure~\ref{fig:power_detuning}(b) shows that $\Delta E \sim 1/\delta$ where delta is the detuning of the pump from the exciton resonance frequency.

In Fig.~\ref{stark}(b) the pump detuning is $\sim 107$ meV, whereas in Fig.~\ref{stark}(e) the pump detuning is $\sim 90$ meV. Considering the linear relationship between the inverse of the detuning and power with the amount of AC Stark shift, the effective pump power for the same shift of $\sim 0.65$ nm for a detuning of 107 meV is 29.7 mW. Hence, the observed enhancement in AC Stark shift when normalized to detuning is $29.7/0.33 = 90$.

\newpage

\subsection{\label{si:fitting} 10. Data processing and fitting method}

\begin{figure}[h]
\includegraphics[width=0.4\columnwidth]{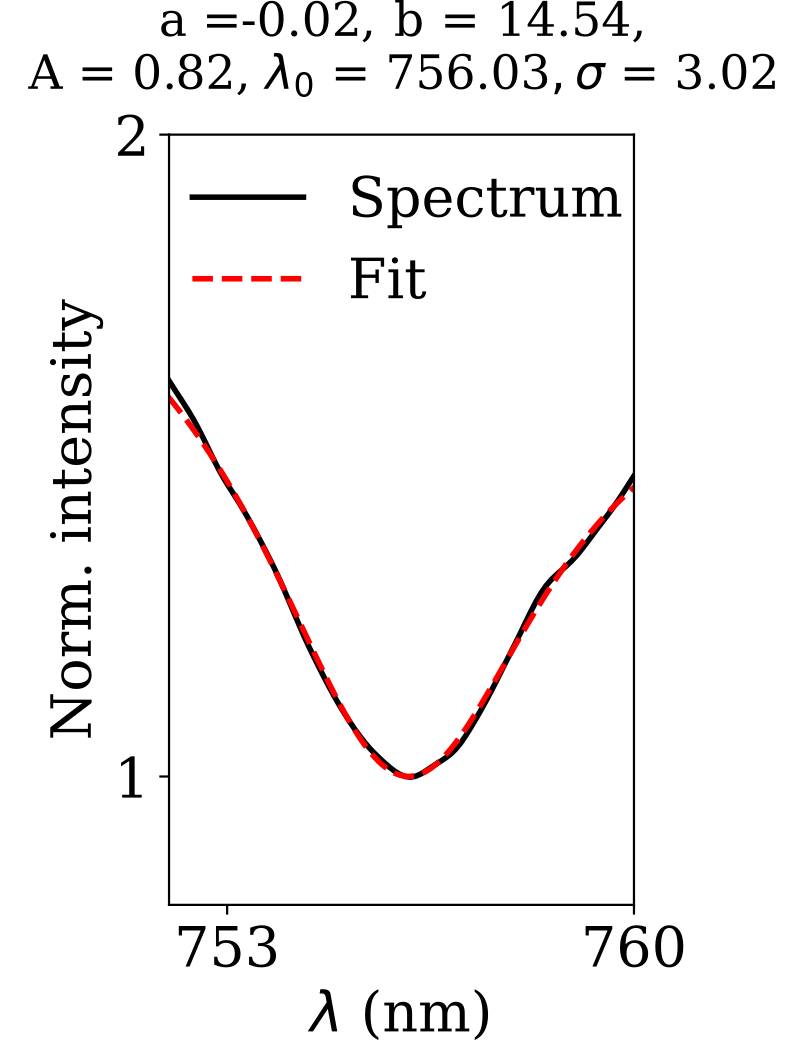}
\caption{\label{fig:fitting} \textbf{Numerical Fitting.} Example of a fitted reflectivity spectrum. The solid black curve is the normalized spectrum, and the red dashed curve is the numerical fit.}
\end{figure}

We use a numerical fitting method to extract the center wavelength and linewidth of the exciton from our measured reflectivity data. First, we normalize each reflectivity spectrum by the minimum reflection intensity at the exciton's resonant wavelength. Next, we truncate the normalized spectrum within a range that captures the exciton's resonance. The truncated normalized spectrum is fitted to the function

\begin{equation*}
    s (\lambda) = \left( a\lambda + b\right) - \dfrac{A \sigma^2}{(\lambda - \lambda_0)^2 + \sigma^2},
\end{equation*}

\noindent where $a\lambda + b$ is a linear function that captures the background spectrum, and $A$, $\lambda_0$ (in nm), $\sigma$ (in nm) are the amplitude, center wavelength, and linewidth of the Lorentzian exciton, respectively. Figure \ref{fig:fitting} shows the example of a fitted spectrum along with the extracted parameters.

\newpage

\subsection{\label{si:power_broadening} 11. Power broadening estimation}

To ascertain if the high peak power of the pump laser might induce power broadening, we perform the following calculations. Let us assume the pump laser is detuned by $\Delta$ and has peak power $P_{\rm{peak}}$. This gives us a Rabi frequency of $\Omega = -\Vec{d} \cdot \Vec{E} / \hbar$, where the electric field $E=\sqrt{\frac{2P_{\rm{peak}}}{\epsilon_0 c A}}$. Here, $\Vec{d}$ is the dipole of the exciton, $\hbar$ is the reduced Planck's constant, $\epsilon_0$ is the vacuum permittivity, $c$ is the speed of light in vacuum, and $A$ is the cross-sectional area of the laser spot. Next, let the linewidth of the exciton be $\Gamma$. Then, the excited state population in the steady-state can be expressed as \cite{steck2007quantum}:

\begin{equation}
    \rho_e = \dfrac{\left( \dfrac{\Omega}{\Gamma} \right)^2}{1 + \left( \dfrac{2\Delta}{\Gamma} \right)^2 + 2 \left( \dfrac{\Omega}{\Gamma} \right)^2}.
\end{equation}

\begin{figure}[h]
\includegraphics[width=0.55\columnwidth]{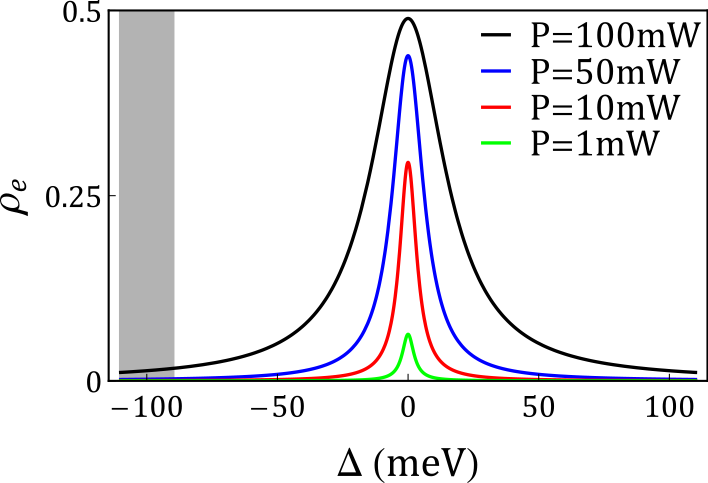}
\caption{\label{fig:power_broad} \textbf{Power broadening estimation.} Excited state population $\rho_e$ for peak intensities corresponding to the average power of $P = $100 mW, 50 mW, 10 mW, and 1 mW respectively. The range of detuning used in our experiments is marked by the grey box.}
\end{figure}

\noindent Assuming linewidth $\Gamma=5$ meV, area $A=1$ $\mu\rm{m}^2$, pulsewidth $\tau_{\rm{pulse}}=150$ fs, repetition rate $f_{\rm{rep}} = 80$ MHz, we calculate the excited state population as a function of pump detuning for peak intensities corresponding to average power of $P = $100 mW, 50 mW, 10 mW, and 1 mW as shown in Fig.~\ref{fig:power_broad}. One must note, that for a pulse width of 150 fs, we fully satisfy the rotating wave approximation (RWA), and are in the adiabatic limit. In our Stark shift experiments we used a detuning of 90-110 meV (marked by grey box). We can observe that $\rho_e$ is negligible at those detunings for the amount of pump power used.

\newpage

\subsection{\label{si:shift_broad} 12. MPP-induced linewidth broadening - AC Stark shift relationship}

\begin{figure}[h]
\includegraphics[width=\columnwidth]{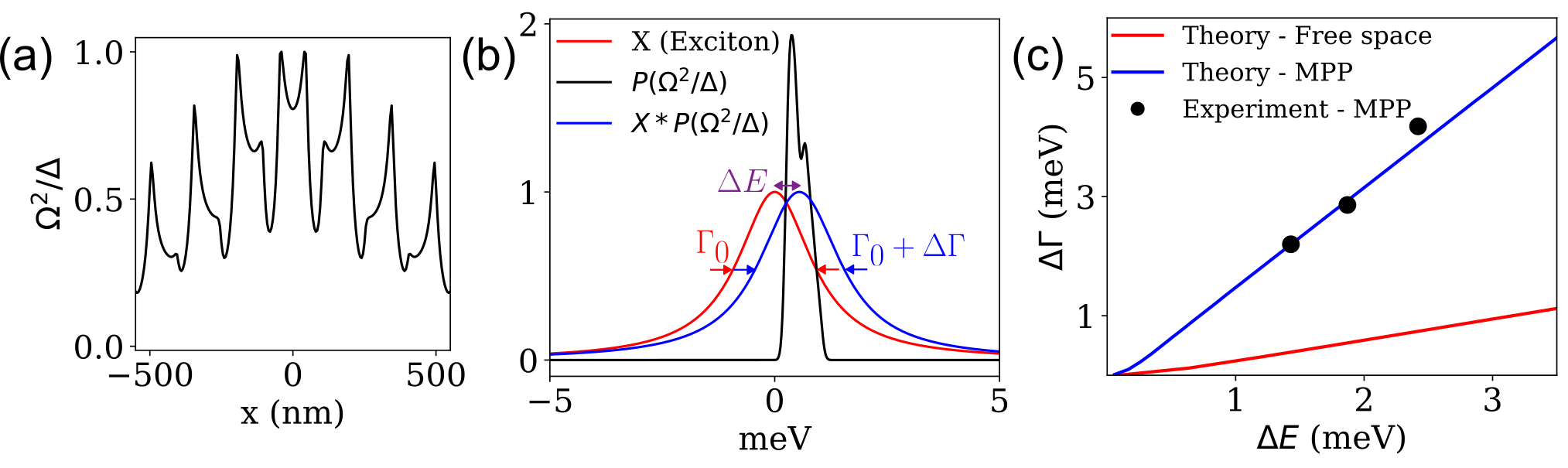}
\caption{\label{fig:shift_broad} \textbf{MPP-induced linewidth broadening - AC Stark shift relationship.} (a) Normalized AC Stark shift ($\sim \Omega^2/\Delta$) as a function of space as experienced by the TMD ML. (b) The probability distribution function of the AC Stark shift (black) shifts the bare exciton X (red) by $\Delta E$ and causes a linewidth broadening of $\Delta \Gamma$ (blue). (c) The blue line shows the theoretical expectation of linewidth broadening corresponding to the AC Stark shift. The black dots represent experimental data by performing local AC Stark shift on MPPs.}
\end{figure}

Utilizing the spatial electric field intensity profile of the MPPs as shown in Fig.~\ref{stark}(a), we estimate the expected linewidth broadening as a function AC Stark shift arising from the sub-diffraction nature of MPPs. Figure~\ref{fig:shift_broad}(a) shows the spatial distribution of the normalized AC Stark shift $\Omega^2/\Delta$. For this spatial distribution, we calculate the probability distribution function (pdf) of the AC Stark shift $P\left( \Omega^2/\Delta \right)$, as shown by the black curve in Fig.~\ref{fig:shift_broad}(b). This pdf modifies the energy of the bare exciton X (red curve) by $\Delta E$ and causes a linewidth broadening of $\Delta \Gamma$. The resultant blue-shifted exciton's spectral distribution function is given by the convolution of X and $P\left( \Omega^2/\Delta \right)$, and shown by the blue curve. By sweeping over laser power, we find the relationship between $\Delta \Gamma$ and $\Delta E$ as shown by the blue line in Fig.~\ref{fig:shift_broad}(c). The black dots represent experimentally measured data by performing local AC Stark shift measurement of excitons on an MPP device, and agree well with our theoretical estimation. The red line shows the relationship between $\Delta \Gamma$ and $\Delta E$ for a free space optical beam with FWHM of 1 $\mu$m. As observed, the free space optical pump induces negligible linewidth broadening compared to its MPP counterpart.

\end{document}